\documentclass[fontsize=12]{article}
\usepackage[a4paper, total={6in, 7.5in}]{geometry}
\usepackage{float}
\usepackage{graphicx}
\usepackage{caption}
\usepackage{subcaption}
\usepackage{url}
\usepackage{multirow}
\usepackage{comment}
\usepackage{amsmath}
\usepackage{authblk}
\usepackage{booktabs} 
\usepackage{xcolor}
\usepackage{braket}
\usepackage{amsmath}

\usepackage{microtype}
\usepackage{breqn}
\usepackage{calrsfs}
\usepackage[ruled,vlined]{algorithm2e}
\usepackage{bbold}

\title{Interbank network reconstruction enforcing density and reciprocity}
\author[1]{Valentina Macchiati\thanks{Corresponding author. Email: valentina.macchiati@sns.it}}
\author[2]{Piero Mazzarisi}
\author[3,4,5]{Diego Garlaschelli}
\affil[1]{Scuola Normale Superiore, Pisa, Italy}
\affil[2]{Università degli Studi di Siena, Siena, Italy}
\affil[3]{IMT School of Advanced Studies, Lucca, Italy}
\affil[4]{Lorentz Institute for Theoretical Physics, University of Leiden, The Netherlands}
\affil[5]{INdAM-GNAMPA Istituto Nazionale di Alta Matematica, Italy}
\date{}

\begin{document}
\maketitle
\begin{abstract}
Networks of financial exposures are the key propagators of risk and distress among banks, but their empirical structure is not publicly available because of confidentiality. This limitation has triggered the development of methods of network reconstruction from partial, aggregate information. 
Unfortunately, even the best methods available fail in replicating the number of directed cycles, which on the other hand play a crucial role in determining graph spectra and hence the degree of network stability and systemic risk.  
Here we address this challenge by exploiting the hypothesis that the statistics of higher-order cycles is strongly constrained by that of the shortest ones, i.e. by the amount of dyads with reciprocated links.
First, we provide a detailed analysis of link reciprocity on the e-MID dataset of Italian banks, finding that correlations between reciprocal links systematically increase with the temporal resolution, typically changing from negative to positive around a timescale of up to 50 days. 
Then, we propose a new network reconstruction method capable of enforcing, only from the knowledge of aggregate interbank assets and liabilities, both a desired sparsity and a desired link reciprocity. We confirm that the addition of reciprocity dramatically improves the prediction of several structural and spectral network properties, including the largest real eigenvalue and the eccentricity of the elliptical distribution of the other eigenvalues in the complex plane.
These results illustrate the importance of correctly addressing the temporal resolution and the resulting level of reciprocity in the reconstruction of financial networks.
\end{abstract}


%\item We observe that reciprocity strongly varies with different aggregation periods, and it cannot be inferred from link density by analyzing empirical interbank market networks (e-MID data). This empirical evidence supports our objective of enhancing F-DCM by constraining not only the link density but also the reciprocity, as the latter provides additional information to the model compared to the former. 

%\item We demonstrate that F-GRM generates an ensemble of networks with spectral properties closer to empirical ones. This is crucial because the spectra of graphs, particularly the leading eigenvalue $\lambda_{max}$, play a key role in terms of systemic stability by accounting for loops and cycle structures that influence the propagation and amplification of an initial shock. 

%\item When assessing system stability, it is important to have a model with stable performance in terms of under/overestimation of the maximum eigenvalue and thus systemic risk. In this regard, F-GRM is a reliable reconstruction model, generating an ensemble of networks with dynamical properties consistently upper-bounding empirical ones and always within the acceptance interval. 

\paragraph{Keywords:}

Financial Networks, Network Reconstruction, Systemic risk, Spectral Properties

\section{Introduction}
Networks serve as a valuable tool for analyzing complex systems, as they clearly illustrate the interactions between components. Often, the structure of these underlying networks is not readily available due to confidentiality, making it challenging to accurately estimate key systemic properties such as resilience to shock propagation. This is particularly true for interbank market networks, where nodes represent financial institutions, links indicate financial ties (loans), and access to bilateral exposures is limited. Instead, we rely on aggregated exposures that are publicly reported in balance sheets.
Systemic risk analysis typically involves reconstructing the underlying network using available information and employing either deterministic or probabilistic approaches. The former yields a unique reconstructed configuration, meaning the first configuration found by the greedy algorithm that is compatible with the given constraints. However, this approach is implicitly biased, as the probability that such a unique configuration is identical to the empirical one is almost zero.
In contrast, the probabilistic approach generates a set of configurations compatible with the available information. Each of these configurations is assigned a probability value, forming an ensemble in statistical mechanics parlance. While the imposed constraints are perfectly matched in deterministic approaches, they are only matched on average in the ensemble of probabilistic ones. Nevertheless, in \cite{gabrielli2024critical}, the authors analyze the problem of network reconstruction using ensemble methods and show that reconstructability is achieved when all the constraints, apart from being replicated on average, are also `sufficiently close' to their expected value in individual typical realizations of the ensemble.

With a focus on financial networks, the review \cite{squartini2018reconstruction} shows that reconstruction methods can be classified depending on the link density of reconstructed configurations, in particular into dense reconstruction methods \cite{di2018assessing,upper2011simulation,wells2004financial}, density-tunable reconstruction models \cite{drehmann2013measuring,mastromatteo2012reconstruction,moussa2011contagion, mazzarisi2017methods}, exact-density methods \cite{cimini2015estimating,cimini2015systemic,mastrandrea2014enhanced,macchiati2022systemic}, and probabilistic approaches such as the copula method for reconstruction \cite{baral2012estimation}, the Bayesian approach \cite{gandy2017bayesian}, a mixed deterministic and stochastic approach leveraging balance sheet information \cite{montagna2017contagion}, a block modeling method \cite{halaj2013assessing}, and the Minimum Density algorithm \cite{anand2015filling}. In \cite{anand2018missing}, the authors access empirical data from 25 markets across 13 jurisdictions and focus on testing the performance of different models \cite{anand2015filling,baral2012estimation,cimini2015systemic,drehmann2013measuring,halaj2013assessing,musmeci2013bootstrapping,upper2011simulation} capable of reconstructing the network based solely on aggregated positions. They demonstrate that the Fitness Induced Directed Configuration model (F-DCM) \cite{cimini2015systemic} outperforms other ensemble methods across various financial markets. In \cite{lebacher2019search}, a similar test is conducted for the payment messages network, concluding that the F-DCM model is a suitable choice for sparse matrices. In the context of the energy trade network, F-DCM continues to emerge as the winner in the horse race among different network reconstruction methods, as highlighted in \cite{xu2023reconstruction}. 
The F-DCM model is recommended when no exogenous information is available in addition to the aggregated exposures, as it performs well on both large sparse networks and small dense networks in terms of edge probabilities and edge values. Given these results, the F-DCM model successfully reconstructs the main structural properties of the empirical network. Furthermore, this model also replicates the dynamic properties \cite{cimini2015systemic}, such as bond percolation properties, the shortest path length distribution, and DebtRank values for synthetic networks. Henceforth, we refer to it as the state-of-the-art.

Graph spectra, particularly the principal eigenvalue $\lambda_{max}$, are pivotal in assessing systemic stability. This importance stems from their ability to capture the loops and cycle structures, which drive the propagation and amplification of an initial shock within the system.
In \cite{neri2020linear}, the authors conduct a linear stability analysis of large dynamical systems on random directed graphs, which are oriented and locally tree-like. They demonstrate that the leading eigenvalue of these random graphs depends solely on a few system parameters, including the mean degree and a parameter characterizing the correlations between in- and out-degrees. Interestingly, they show that dynamical systems on such directed graphs can remain stable even when the degree distribution has unbounded support. In contrast, in the undirected case, such a system becomes unstable if the system size is sufficiently large. The undirected case and the locally tree-like structure can be viewed as extremes in terms of stability and reciprocity structure. In the former, all links are bidirected, while in the latter, there are no bilateral links. When considering not just the leading eigenvalue but the entire spectrum, we observe that it lies in the real domain for the undirected case and in the complex domain for the directed case. In \cite{sommers1988spectrum}, it is shown that the average eigenvalue distribution of square real random non-symmetric matrices (directed case) is uniform in an ellipse, with real and imaginary axes dependent on the reciprocity value. When all links are bidirected (undirected case), Wigner’s semicircle law is recovered. In \cite{squartini2013early}, the authors analyze the occurrences of dyadic motifs, i.e. the number of reciprocated, non-reciprocated, and empty dyads, in the quarterly Dutch interbank network from 1998 to 2008. They compute the z-score to quantify the difference between these occurrences in the empirical network and those generated by random null models. When the considered null model is the directed random graph (DRG), the DRG consistently underestimates the empirical reciprocity. In contrast, the directed configuration model (DCM) initially underestimates it from 1998 to 2004, albeit performing better than the DRG, and then begins to overestimate it until the network configuration collapses in 2008. 

In the banking system, reciprocity corresponds to the presence of credits and loans between counterparties within a given period. The main reason banks enter the deposit market is to extend loans to each other to match the capital requirements imposed by regulation day by day. As such, banks can behave as lenders and borrowers depending on their monetary needs. There are two main reasons for increasing the chances of observing credits and loans between counterparties simultaneously. First, the typical maturity of credit is one day (overnight market), one week, or one month, like in the Italian electronic Market of Interbank Deposit (e-MID) analyzed below. Moreover, the loan terms are defined at the moment of credit extension and cannot be changed hereafter. As such, a credit and a loan between two counterparties ``overlapping'' over some period can be described as a pair of reciprocated links in the interbank network. Second, preferential lending (i.e. the tendency of extending loans with preferential counterparties) is recognized as a key aspect of a credit market; e.g. see \cite{hatzopoulos2015quantifying} for a quantitative analysis of the e-MID market. It also represents a (statistically significant) explanation for the stability (i.e. persistence) of links (credits) in both directions for the e-MID market, see \cite{mazzarisi2020dynamic}. All these aspects justify the inclusion of reciprocity as a key feature of the interbank network, in particular for large aggregation scales within which preferential lending plays a major role. Finally,
the presence of ``credit cycles'' is crucial in terms of systemic risk: a missed payment of a bank to a counterparty can induce financial distress, which feeds back to the first bank in the case of bilateral credit exposures. The presence of network cycles has been recognized as crucial for risk propagation, e.g. see \cite{bardoscia2017pathways}, and its impact on systemic risk is related to the largest eigenvalue of the adjacency matrix \cite{bardoscia2015debtrank}.

Given the relevance of reciprocity in both spectral distribution and system stability, we would like to concentrate on it and our case study is the interbank market network.
In the scientific literature, there exist models capturing reciprocity patterns in networks as opposed to the F-DCM. However, they are not devised for reconstruction, typically. For example, the Reciprocal Configuration Model (RCM) describes all possible dyadic configurations at the node level, distinguishing both outward and inward links, and if they are reciprocated or not. Also, DCM has been generalized to account for reciprocity at the network level, i.e. globally in \cite{engel2019reconstructing}. Both of them require, however, the degree sequence as an input of the method. In general, such information is not available for interbank networks due to data confidentiality. Our goal is to propose an extension of the F-DCM that accounts for link reciprocity at the network level, thus representing one more parameter only to be tuned to solve the reconstruction problem. Interestingly, similar to network density, the empirical value of reciprocity is generally available in the literature; see, e.g., \cite{bargigli2015multiplex,brandi2018epidemics,roukny2014network}.

The remainder of the paper is structured as follows: Section 2 provides a brief overview of the exponential random graph models and introduces our model. Section 3 describes the dataset we use to corroborate our findings. Section 4 presents the results. Section 5 is for final remarks and conclusions. Appendix sections contain supplementary material supporting the results.

\section{Methods}

%{\color{red} 
Within the probabilistic framework, network reconstruction methods can be classified into three groups depending on the particular estimation approach adopted for applications. Bayesian methods rely on probabilistic models to estimate network parameters, incorporating prior knowledge and updating beliefs based on observed data. These methods are robust in handling uncertainty but can be computationally intensive; see, e.g., \cite{gandy2017bayesian,peixoto2018reconstructing}. Statistical approaches, such as correlation or regression analysis, identify relationships between variables to reconstruct networks. They are generally simpler and faster but may struggle with capturing non-linear interactions or dependencies. When complemented with machine learning techniques, such drawbacks can be overcome to achieve better performances. For example, compressed sensing, a recently developed paradigm in convex optimization, is suggested in \cite{wang2011network} to recover sparse signals or structures from incomplete data. LASSO \cite{han2015robust} and Signal LASSO \cite{shi2021inferring} extend this method by incorporating regularization to promote sparsity, helping to identify the most significant connections in complex networks. Adaptive Signal LASSO \cite{shi2023robust} further enhances this capability by dynamically adjusting penalties based on data. In the same context, in order to overcome the issue of hidden nodes, a robust two-stage reconstruction method is proposed in \cite{deng2022two} to infer the complete topology from available time series data of accessible nodes. Moreover, deep learning techniques \cite{liu2023si} can enhance the inference of network structure using least squares generative adversarial networks. In \cite{huang2024one}, e.g., fuzzy neural networks and a predictive model are used to adapt and control complex industrial processes under varying conditions. These methods can handle complex and high-dimensional data, offering high accuracy and scalability, but they often require extensive training data and computational resources.

Finally, in the general context of statistical methods, the maximum entropy approach \cite{squartini2018reconstruction,park2004statistical} represents the best compromise to achieve high flexibility, scalability, and control in reconstructing networks from limited information. Relying on the point estimation of maximum entropy distributions, it derives the least biased probability distribution consistent with the limited information encoded in several constraints, representing the available information about the network to be reconstructed. Its effectiveness largely depends on the available network data and whether local or global constraints are used. Local constraints are specific to nodes, such as the degree sequence or strengths, while global constraints pertain to the overall system, like link density or reciprocity. It is clear that the more information is available, the better the reconstruction will be. In this paper, we propose a new network reconstruction method within the general context of the maximum entropy approach by using partial information on global density and reciprocity, together with aggregated banks' exposures, for reconstructing the interbank network.
%}

\paragraph{Matrix representation}
The weighted adjacency  $W=\{w_{ij}\}$ is the $N\times N$ matrix representation of a network with $N$ nodes, where the generic element $w_{ij}$ denotes the weight of the link from node $i$ to node $j$. The adjacency matrix $A=\{a_{ij}\}$ represents the binary version of $W$, with generic element $a_{ij} = 1$ if $w_{ij} > 0$ (and $0$ otherwise). We define in/out-degree and strengths of the node $i$: $k_i^{in}=\sum_j a_{ji}$, $k_i^{out}=\sum_j a_{ij}$, $s_i^{in}=\sum_j w_{ji}$, $s_i^{out}=\sum_j w_{ij}$, respectively. 
\\\\
\textbf{Exponential Random Graphs (ERGs)} \cite{park2004statistical,wasserman1994social,robins2007introduction} are characterized as the ensemble of graphs where the probability $P(G)$ is determined by two distinct optimization processes. The first process, entropy maximization, ensures that the derived probability distribution only encodes information from the selected constraints. $P(G)$ is the probability associated with the graph $G$ in the ensemble $\mathcal{G}$. This probability is chosen by maximizing the Shannon-Gibbs entropy $S$ 
\begin{equation}
    S=-\sum_{G \in \mathcal{G}} P(G)\ln P(G)
\end{equation}
such that the expectation value of the observables $\{\langle C_i(G)\rangle_{\mathcal{G}}\}$ are equal to the observed values $\{C_i^*\}$
\begin{equation}
    \sum_{G \in \mathcal{G}} P(G)C_i(G)=C_i^*~,~ \sum_{G \in \mathcal{G}} P(G)=1 .
\end{equation}
By introducing the Lagrange multipliers $\nu$, {$\theta_i$} the maximum entropy probability distribution is obtained by:
\begin{equation}
   \frac{\partial}{\partial P(G)} \biggl\{ S+\nu \Bigl(1-\sum_{G \in \mathcal{G}} P(G) \Bigr) +\sum_i \theta_i \Bigl(C_i^*- \sum_{G \in \mathcal{G}} P(G)C_i(G) \Bigr) \biggr\}=0
\end{equation}
The solution is then:
\begin{equation}
    P(G|\theta)=\frac{e^{-H(G,\theta)}}{Z(\theta)}
\end{equation}
where $H(G,\theta)=\sum_i \theta_i C_i$ is the graph Hamiltonian and $Z(\theta)$ is the partition function which properly normalizes the probability distribution. 
The second process, likelihood maximization, ensures that the value of the imposed constraints aligns with the observed value without any statistical bias. The value Lagrange multipliers $\theta$s are obtained by log-likelihood maximization. 

When the imposed constraints are the in and out degree sequences, the corresponding ERG model is called \textit{Directed Configuration Model} (DCM).
The DCM Hamiltonian and the link probability are, respectively,
\begin{equation}
    H_{DCM}=\sum_{i=1}^N \alpha_i k_i^{out}+\beta_i k_i^{in}=\sum_{i=1}^N \sum_{j\neq i=1}^N (\alpha_i + \beta_j) a_{ij} ,
\end{equation}
\begin{equation}
    p_{ij}^{DCM}=\frac{x_iy_j}{1+x_iy_j} ,
\label{pij_DCM}
\end{equation}
where $x_i=e^{-\alpha_i}$, $y_i=e^{-\beta_i}$ are the exponential of the Lagrange multipliers that are associated with the out- and in-degree, respectively.
In the case of DCM, there are 2N Lagrange multipliers [$\Vec{x}$, $\Vec{y}$] to be tuned or estimated.
\begin{comment}
\begin{equation}
    P(A|\theta)=\Pi_i\Pi_{j\neq i} p_{ij}^{a_{ij}} (1-p_{ij})^{a_{ij}}
\end{equation}
where the link probability is $p_{ij}=\frac{x_iy_j}{1+x_iy_j}$ and $x_i=e^{-alpha_i}$, $y_i=e^{-beta_i}$.
\end{comment}

\subsection{Fitness-Directed Configuration Model}
However, the use of the Directed Configuration Model is not feasible when the degrees of nodes are unknown, a situation that often arises due to confidentiality or data scarcity. This issue can be addressed by employing the
\textit{fitness} ansatz \cite{caldarelli2002scale}, which posits that the connection probability between any two nodes is determined by peculiar non-topological properties of the involved nodes. More specifically, it is postulated that the ‘activity’ of each node $i$ in the network is encapsulated by an inherent quantity known as fitness, which is likely linked to the Lagrange multipliers $x_i,y_i$ that control that node’s out- and in-degree through a monotone functional relationship. 

In this paper, the focus is on the interbank market network whose nodes are banks and links denote lending relationships between them. Due to confidentiality, the only local data available are the total interbank assets/liabilities from the public balance sheets of institutions (strengths), but the empirical values of a few global topological metrics can be found in the literature \cite{bargigli2015multiplex}. In this context, the \textit{fitness} ansatz has been successfully validated in previous studies \cite{cimini2015systemic,squartini2018reconstruction,mazzarisi2017methods} by observing a strong linear correlation between the Lagrange multipliers of nodes’ degrees and the total assets ($\mathcal{A}$) and liabilities ($\mathcal{L}$) values of the corresponding banks: $x_i\equiv \sqrt{b} \mathcal{A}_i$ and $y_i\equiv \sqrt{c} \mathcal{L}_i$. In the \textit{Fitness induced Directed Configuration Model (F-DCM)} the link probability in eq. \ref{pij_DCM} is transformed as follows:
\begin{equation}
    p_{ij}^{F-DCM}=\frac{z\mathcal{A}_i\mathcal{L}_j}{1+z\mathcal{A}_i\mathcal{L}_j}
\label{eq:pij_cimi}
\end{equation}
where $z=\sqrt{bc}$ is the free parameter that is tuned by imposing the link density, denoted as $d$. This parameter $z$ is found by solving the following nonlinear equation:
\begin{equation}
    \sum_{i,j\neq i}p_{ij}^{F-DCM}=N(N-1)d
    \label{eq:sysFDCM}
\end{equation}
where $d$ is the link density of the network, namely the total number of links divided by all the possible pairs of distinct nodes.
As aforementioned, this model is considered state-of-the-art for interbank market networks. In \cite{anand2018missing}, the performance of various reconstruction methods was evaluated using empirical bilateral data from 25 markets across 13 jurisdictions. These markets included interbank networks, payment networks, networks of repurchase agreements, foreign exchange derivatives, credit default swaps, and equities. F-DCM \cite{cimini2015systemic} demonstrated superior performance and was identified as ``the clear winner among ensemble methods.''
\subsection{Encoding the loop structure}
An effective reconstruction method should aim at replicating as many topological properties of the real network as possible, while still requiring as input as little aggregate empirical information as possible, given what can be known and observed about the real system. Loop structures are indeed crucial as they determine whether an initial shock is propagated and amplified. However, the \textit{Directed Configuration Model} does not consider any information on cycle structure, unlike the \textit{Reciprocal Configuration Model} (RCM) and the \textit{Global Reciprocity Model} (GRM) \cite{picciolo2012role}. All three models constrain the degree sequences, but the GRM and RCM also constrain the number of loops of order two; the GRM does this globally, while the RCM does it locally at the node level.

\textit{Global Reciprocity Model} constraints the in/out-degree and the global number of bidirected links $L^\leftrightarrow=\sum_{i\neq j}a_{ij}a_{ji}$.The GRM Hamiltonian and the link probability are:
\begin{equation}
    H_{GRM}=\sum_{i=1}^N \sum_{j\neq i=1}^N (\alpha_i + \beta_j) a_{ij} + \gamma\sum_{i\neq j}a_{ij}a_{ji},
\label{eq:HGRM}
\end{equation}
\begin{equation}
    p_{ij}^{GRM \rightarrow}= \frac{x_iy_j}{1+x_iy_j+x_jy_i+z^2x_iy_jx_jy_i} ,
\end{equation}
\begin{equation}
    p_{ij}^{GRM \leftrightarrow}=\frac{z^2x_iy_jx_jy_i}{1+x_iy_j+x_jy_i+z^2x_iy_jx_jy_i}
\end{equation}
where $x_i=e^{-\alpha_i}$, $y_i=e^{-\beta_i}$, $z=e^{-\gamma}$ are the exponential of the Lagrange multipliers that are associated with the out-degree, in-degree and number of cycle of order two, respectively. In the case of GRM, there are 2N+1 Lagrange multipliers [$\Vec{x}$, $\Vec{y}$, $z$] to be tuned or estimated.

\textit{Reciprocal Configuration Model} constraints separately the non-reciprocated out-degree sequence $k_i^\rightarrow=\sum_{j\neq i}a_{ij}(1-a_{ji})$, the non-reciprocated in-degree sequence $k_i^\leftarrow=\sum_{j\neq i}a_{ji}(1-a_{ij})$ and the reciprocated degree sequence $k_i^{\leftrightarrow}=\sum_{j\neq i}a_{ij}a_{ji}$. 
The RCM Hamiltonian and the mono- and bi-directed link probability are:
\begin{equation}
    H_{RCM}=\sum_{i=1}^N \alpha_i k_i^{\rightarrow}+\beta_i k_i^{\leftarrow} + \gamma_i k_i^{\leftrightarrow} ,
\end{equation}
\begin{equation}
    p_{ij}^{RCM \rightarrow}= \frac{x_iy_j}{1+x_iy_j+x_jy_i+z_iz_j} ,
\end{equation}
\begin{equation}
    p_{ij}^{RCM \leftrightarrow}=\frac{z_iz_j}{1+x_iy_j+x_jy_i+z_iz_j}
\end{equation}
where $x_i=e^{-\alpha_i}$, $y_i=e^{-\beta_i}$, $z_i=e^{-\gamma_i}$ are the exponential of the Lagrange multipliers that are associated with the non-reciprocated out-degree, the non-reciprocated in-degree and the reciprocated degree, respectively. In the case of GRM, there are $3N$ Lagrange multipliers [$\Vec{x}$, $\Vec{y}$, $\Vec{z}$] to be tuned or estimated.
\subsection{Fitness-Global Reciprocity Model}
In the context of interbank networks, the only available local data are the total interbank assets and liabilities. As previously mentioned, the Fitness-induced Directed Configuration Model \cite{cimini2015systemic} requires a local constraint (the strengths) and a global constraint (the number of connections). This paper seeks to enhance this state-of-the-art model to achieve a cycle structure more akin to the empirical one. To this end, we introduce an additional constraint: the number of bidirected links. This is a global property, the value of which can be found in the literature. While we could have imposed other constraints, particularly local ones, they are unfortunately not available. By imposing the empirical number of cycles of order two, we expect to better capture the structure of higher-order cycles. 

As F-DCM follows from the directed configuration model by imposing the fitness ansatz, our model the \textit{Fitness induced Global Reciprocity Model} (F-GRM), is derived from the global reciprocity model by imposing the same fitness ansatz,  $x_i\equiv \sqrt{b} \mathcal{A}_i$ and $y_i\equiv \sqrt{c} \mathcal{L}_i$. The GRM Hamiltonian is given in eq. \ref{eq:HGRM}. By introducing the fitness ansatz, the probabilities of mono- and bi-directed links are transformed as follows:

\begin{equation}
p_{ij}^{F-GRM \rightarrow}=\frac{u \mathcal{A}_i \mathcal{L}_j}{1+u \mathcal{A}_i \mathcal{L}_j+u \mathcal{A}_j \mathcal{L}_i+u^2 v^2 \mathcal{A}_i \mathcal{L}_j \mathcal{A}_j  \mathcal{L}_i}
\label{eq:pmono_grm}
\end{equation}
\begin{equation}
p_{ij}^{F-GRM \leftrightarrow}=\frac{u^2 v^2  \mathcal{A}_i \mathcal{L}_j \mathcal{A}_j  \mathcal{L}_i}{1+u \mathcal{A}_i \mathcal{L}_j+u \mathcal{A}_j \mathcal{L}_i+u^2v^2  \mathcal{A}_i \mathcal{L}_j \mathcal{A}_j  \mathcal{L}_i}
\label{eq:pbi_grm}
\end{equation}
where $u=\sqrt{bc}$, $v=e^{-\gamma}$  are the free parameters that are tuned by imposing that the observed values for link density $d$ and reciprocity $r$ (namely the number of links in both directions divided by the number of total links) match their expected values according to the model. The generic probability of a link between nodes $i$ and $j$ is
\begin{equation}
p_{ij}^{F-GRM}=p_{ij}^{F-GRM \rightarrow}+p_{ij}^{F-GRM \leftrightarrow}.
\label{eq:p_grm}
\end{equation}
The parameters $u, v$ are estimated by solving the following system of nonlinear equations\footnote{Further details on the parameters' estimation procedure could be found in Appendix \ref{appendix:trf}.}:
\begin{equation}
\left\{
\begin{aligned}
    \sum_{i,j\neq i}p_{ij}^{F-GRM}&=N(N-1)d\\\
    \frac{\sum_{i,j\neq i}p_{ij}^{F-GRM \leftrightarrow}}{\sum_{i,j\neq i}p_{ij}^{F-GRM }}&=r
\end{aligned}
\label{eq:sysFGRM}
\right.
\end{equation}
where $d$ and $r$ are the observed values for link density and reciprocity.

\section{e-MID Data}
The electronic Market for Interbank Deposits (e-MID) is a trading platform for unsecured money-market loans, accessible to both Italian and foreign banks. We have access to the interbank transactions finalized on e-MID\footnote{Further details are available in Appendix \ref{appendix:data}.} from January 1999 to December 2014. For each contract, we have access to the amount exchanged, the date, the IDs of the lender and the borrower banks, and the contract maturity.  

Our analysis is limited to transactions among Italian banks, not only because they constitute the majority both in number (98\%) and in volume (85\%) in 2011, but also due to their relatively minor fluctuations in terms of active nodes over the years.
In Fig. \ref{Fig:enne}, we observe a gradual decrease in the number of active banks over the years. This decline has remained consistently stable over the years.
While previous literature extensively explores the evolution of the e-MID network across different years and aggregation periods, the focus often centres on the emergence of preferential lending relationships rather than a detailed examination of the relationship between density and reciprocity.
In \cite{finger2013network}, authors assert that a quarterly aggregation provides the best trade-off, capturing both the emergence of preferential relationships and the dynamic evolution of the system. Daily aggregation tends to exhibit a less informative random structure, while yearly aggregation might be problematic due to the potential rapid evolution of the banking network, especially during unstable times.
In \cite{fricke2015core}, observations indicate that the reduction in interbank lending during the 2008 financial crisis primarily resulted from the activity reduction of core banks. The e-MID overnight market is also influenced by the ECB's unconventional measures, such as long-term refinancing operations (LTROs) at the beginning of 2012.
In \cite{barucca2018organization}, findings illustrate that under normal conditions, the most likely network structure is bipartite. However, following the LTRO, the network adopts a random organization. Moreover, depending on the granularity of the data, other structures than core-periphery can better fit the data \cite{barucca2016disentangling}

\subsection{Description of empirical networks}
e-MID and interbank markets in general can be easily represented as a directed network, where interbank loans constitute the direct exposures between banks and allow for the propagation of financial distress in the system. The generic entry $a_{ij}$ of the adjacency matrix $A=\{a_{ij}\}_{i,j=1,\ldots,N}$ takes value $a_{ij}=1$ if there is at least one lending relation between bank $i$ (lender) and $j$ (borrower) in the aggregation period under investigation, otherwise $a_{ij}=0$. Self-loops, namely links that connect nodes with themselves, are not admitted, meaning that $a_{ii}=0$.

In this work, we introduce a new reconstruction model F-GRM that goes beyond F-DCM by incorporating not only the link density ($d$) but also the link reciprocity ($r$). We begin by analyzing the relationship between density and reciprocity across various aggregation periods and different years. This is essential to determine under which conditions link reciprocity can be reproduced by simpler reconstruction models, such as the directed random model and the F-DCM, and when a more complex method, like ours, becomes necessary. It is
\begin{equation}
    d=\frac{L}{N(N-1)}=\frac{\sum_{i \neq j} a_{ij}}{N(N-1)},
    \label{eq:dens}
\end{equation}
\begin{equation}
    r=\frac{L^{\leftrightarrow}}{L}=\frac{\sum_{i \neq j} a_{ij}a_{ji}}{\sum_{i \neq j} a_{ij}},
    \label{eq:rec}
\end{equation}
where $L$ is the number of links, $L^{\leftrightarrow}$ is the number of bilateral links, $A=\{a_{ij}\}$ is the adjacency matrix and $N$ is the number of active nodes.

\begin{figure}[ht]
     \centering
     \begin{subfigure}[b]{0.495\textwidth}
         \centering
         \includegraphics[width=\textwidth]{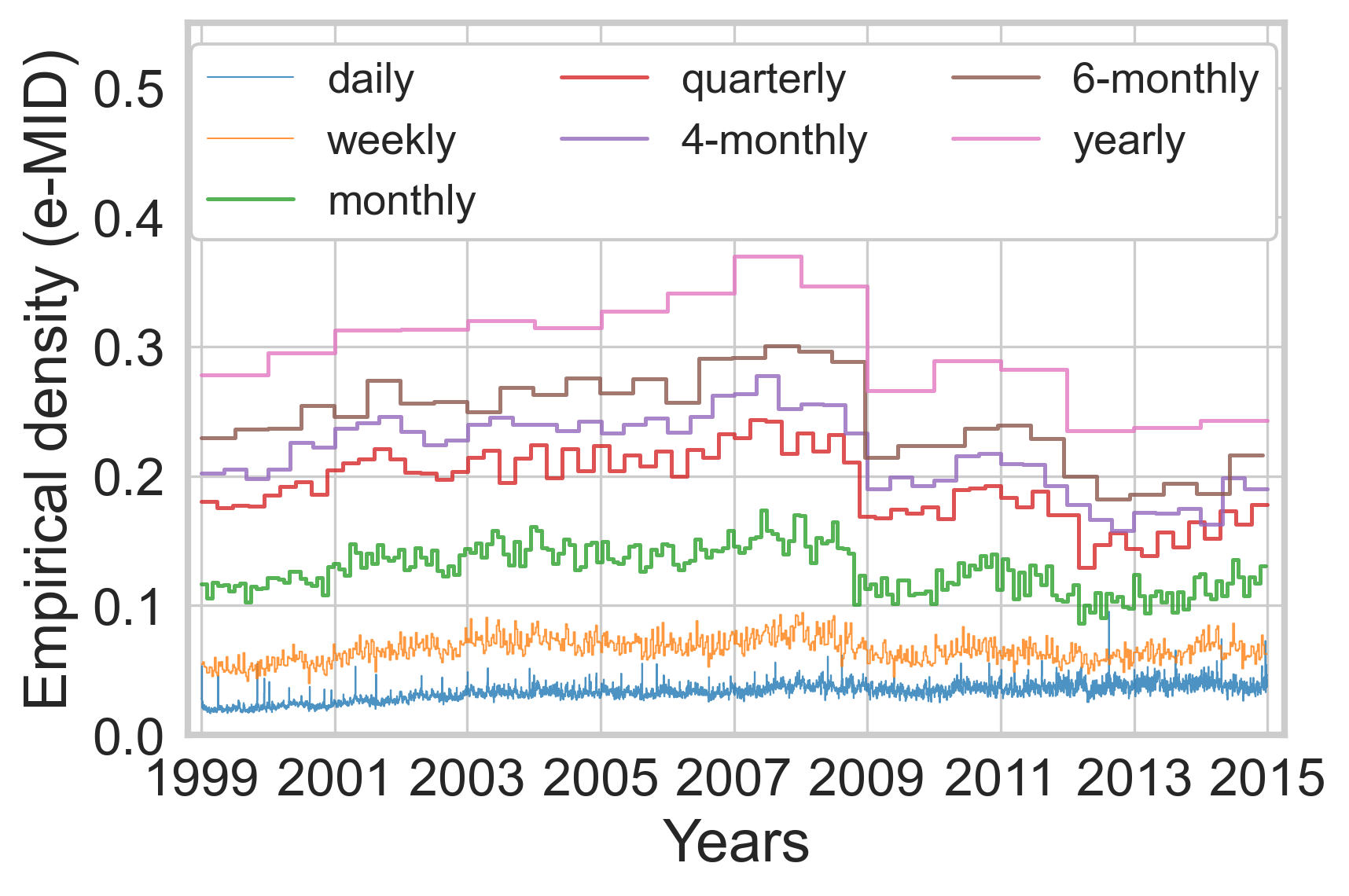}
         \caption{Density $d$}
         \label{Fig:density}
     \end{subfigure}
     %\hfill
     \begin{subfigure}[b]{0.495\textwidth}
         \centering
         \includegraphics[width=\textwidth]{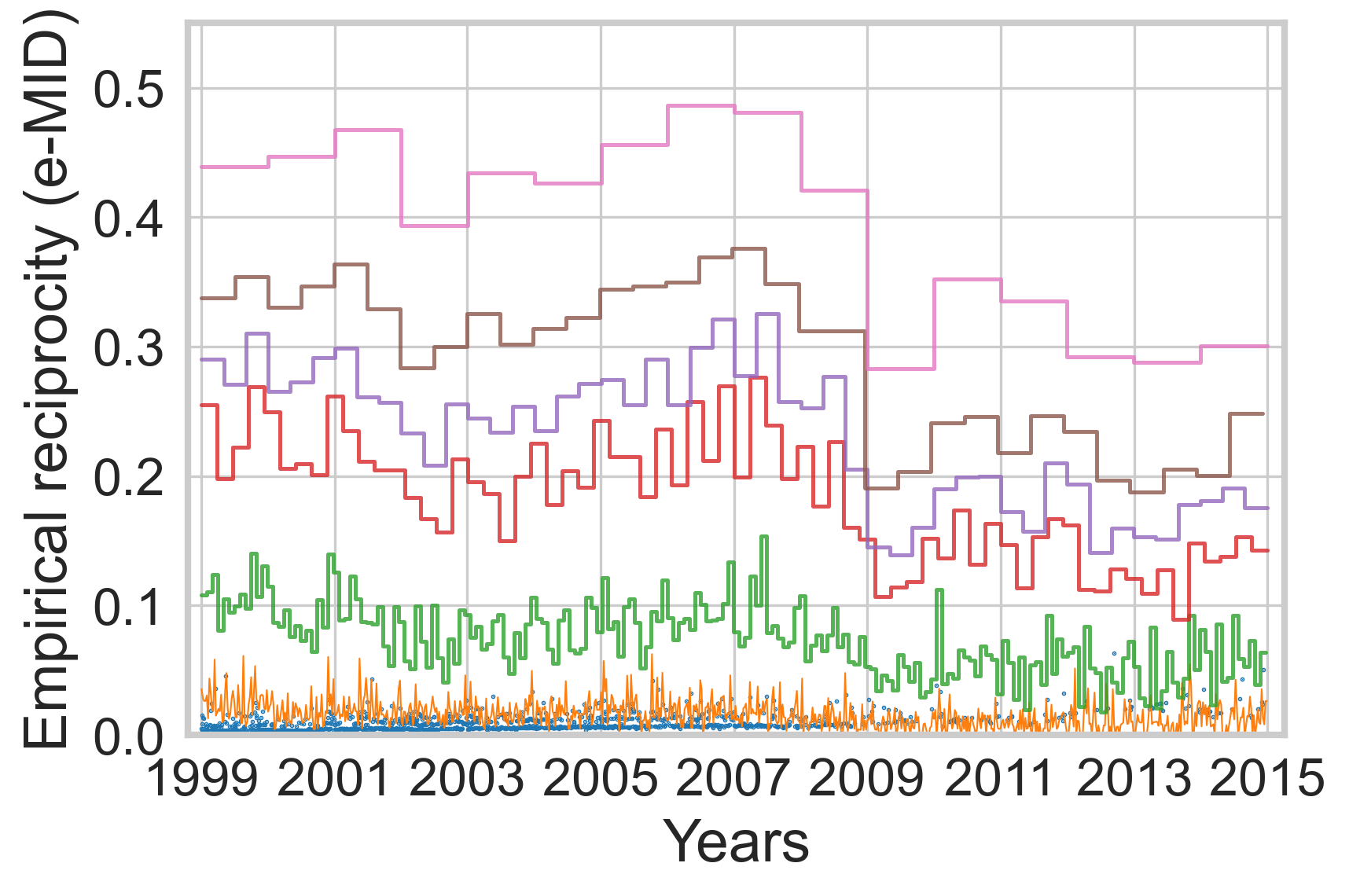}
         \caption{Reciprocity $r$}
         \label{Fig:reciprocity}
     \end{subfigure}
     %\hfill
     \begin{subfigure}[b]{0.495\textwidth}
         \centering
         \includegraphics[width=\textwidth]{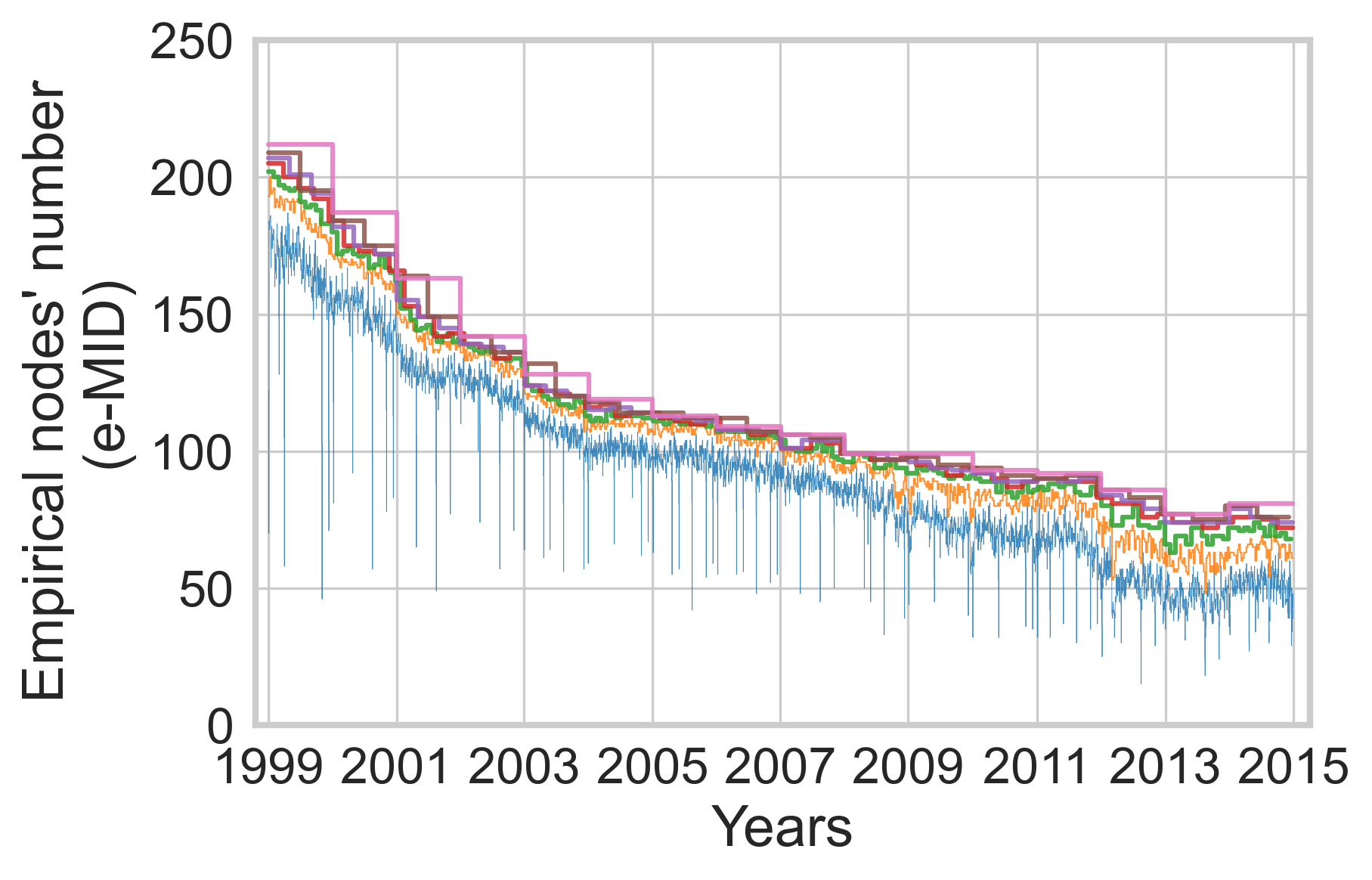}
         \caption{Number of active nodes $N$}
         \label{Fig:enne}
     \end{subfigure}
\caption{Empirical density, reciprocity and number of active nodes in the period 1999-2014. Different colors correspond to different aggregation periods.}
\label{Fig:dens_rec}
\end{figure}

In Figs. \ref{Fig:density} and \ref{Fig:reciprocity}, we investigate the evolution of density and reciprocity across different aggregation periods. For directed random networks, the link probability is $p$ and the link density and reciprocity have the same value $d=r=p$. Using the random case as the benchmark, we observe that daily (yearly) empirical networks are under- (over-) reciprocated, while quarterly networks are comparable. This implies that in the quarterly case, a random model that only imposes link density can also capture the reciprocity value. Additionally, both density and reciprocity exhibit less stability as the aggregation period increases. Specifically, a notable decline in density and reciprocity values has emerged from early 2009, contrasting with the relatively stable decrease in the number of active nodes depicted in Fig. \ref{Fig:enne}.
Upon analyzing the evolution of density and reciprocity values from daily to yearly aggregations, we note that density increases at a slower rate than reciprocity. This suggests a higher likelihood of a borrower becoming a lender, or vice versa, as the aggregation period increases, compared to observing a new lending relationship. This observation supports the presence of preferential lending, potentially linked to the cost of ``trusting'' new counterparties. Similar behaviour is observed in other economic networks, such as the World Trade Network \cite{garlaschelli2005structure}. Fig. \ref{Fig:dens_rec} illustrates that density values cannot be used to infer reciprocity due to the non-trivial relation between them, depending in particular on either the aggregation scale or the year of analysis.
In conclusion, this empirical evidence supports our objective of advancing the F-DCM by imposing not only link density but also reciprocity. Indeed, the latter provides additional information to the model compared to the former.

\section{Results}
This paper aims to enhance F-DCM to generate a network ensemble with spectra and loop structures that better reproduce empirical observations. While the F-DCM model ensures the correct number of links ($L$), in our model, we also constrain the number of bilateral links ($L^{\leftrightarrow}$). The scarcity of data due to confidentiality constraints prevents the use of local constraints to achieve a more refined ensemble. Thus, we explore how imposing only the empirical value of global link reciprocity allows us to obtain spectra that closely align with the empirical ones.
The link reciprocity of our model is, by definition, equal to the constrained empirical reciprocity ($r_{F\text{-}GRM}=r_{emp}$), while in the case of F-DCM, it is directly derived by the constrained link density, which is 
\begin{equation}
    r_{F\text{-}DCM}=\frac{\sum_{i \neq j} p_{ij}^{F\text{-}DCM}p_{ji}^{F\text{-}DCM}}{\sum_{i \neq j} p_{ij}^{F\text{-}DCM}}
\end{equation}
where $p_{ij}^{F\text{-}DCM}$ is given by Eq. \ref{eq:pij_cimi}.
\subsection{Empirical and expected reciprocity}
\begin{figure}[h!]
    \centering
    \includegraphics[width=0.8\textwidth]{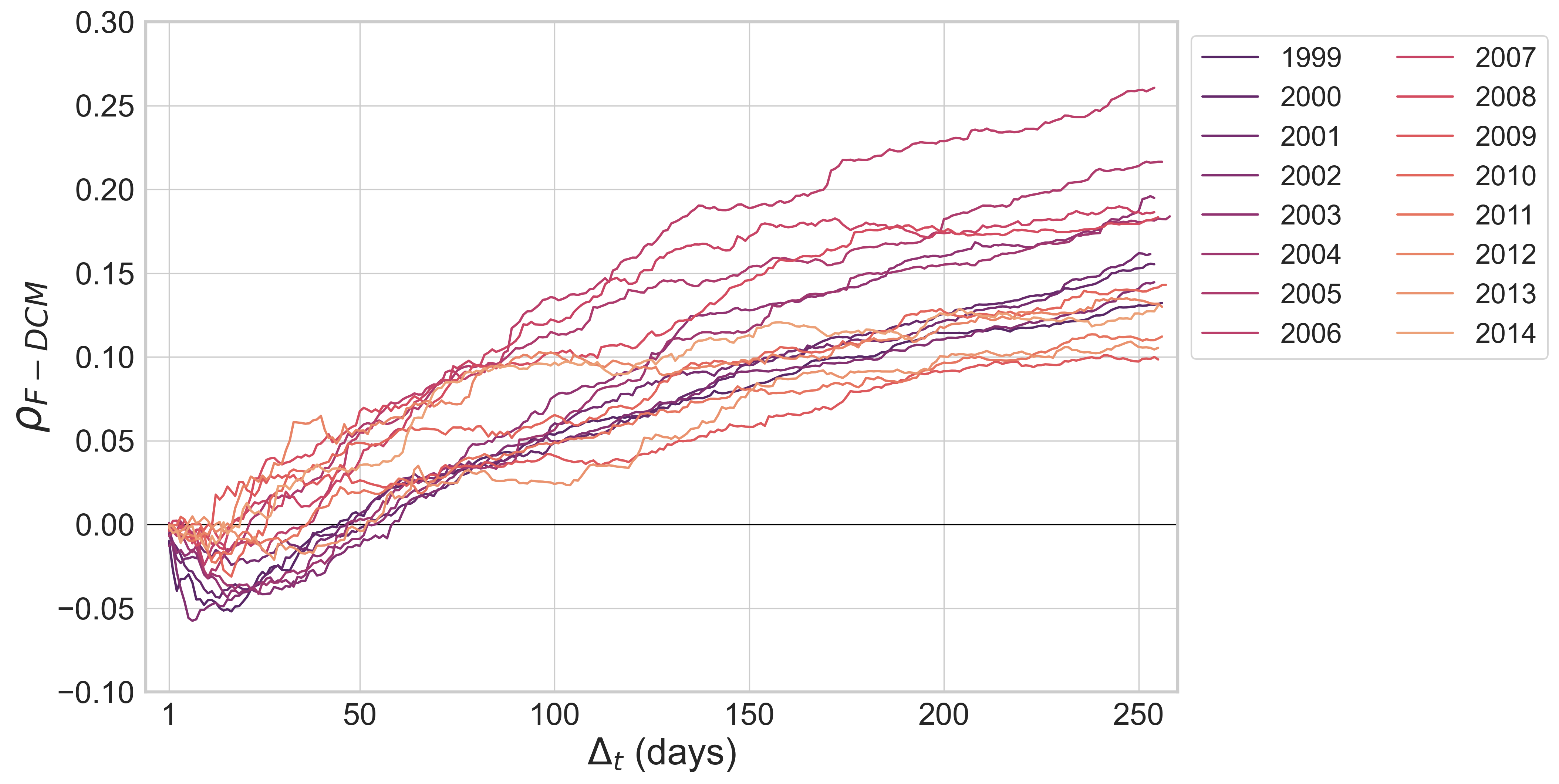}
     \caption{$\rho_{F\text{-}DCM}$ when we consider different aggregation periods $\Delta_t$ (daily unit) in the period 1999-2014. Different colors correspond to different years.}
    \label{Fig:rho_cimi}
\end{figure}

First, we analyse which aggregation period the F-DCM replicates the empirical link reciprocity in. It is worth noting that F-GRM predicts $r_{emp}=r_{F\text{-}GRM}$ by construction. To quantify the difference between the empirical reciprocity $r_{emp}$ and the expected reciprocity by F-DCM $r_{F\text{-}DCM}$, we define the following variable:
\begin{equation}
    \rho_{F\text{-}DCM}=\frac{r_{emp}-r_{F\text{-}DCM}}{1-r_{F\text{-}DCM}}.
\end{equation}
When $\rho_{F\text{-}DCM}$ approaches zero, F-DCM aligns with the empirical reciprocity. However, for positive (negative) values, the ensemble of networks generated by the F-DCM model tends to under (over) estimate the empirical reciprocity. This is illustrated in Fig. \ref{Fig:dens_rec}. The topological properties of the e-MID network exhibit instability over the considered period. We initially segment the dataset into different years and subsequently examine all possible aggregation periods $\Delta_t$ (in days) ranging from daily to yearly. In Fig. \ref{Fig:rho_cimi}, a consistent pattern emerges over the years: at the daily level, $\rho_{F\text{-}DCM}$ is negative but close to zero, then decreases, reaches a minimum, crosses zero, turns positive, and peaks at the yearly aggregation level.

\begin{figure}[h]
     \centering
     \begin{subfigure}[b]{0.45\textwidth}
         \centering
         \includegraphics[width=\textwidth]{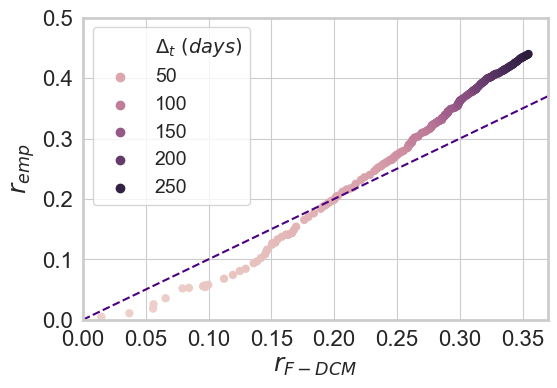}
         \caption{1999}
         \label{Fig:rec_1999}
     \end{subfigure}
     %\hfill
     \begin{subfigure}[b]{0.45\textwidth}
         \centering
         \includegraphics[width=\textwidth]{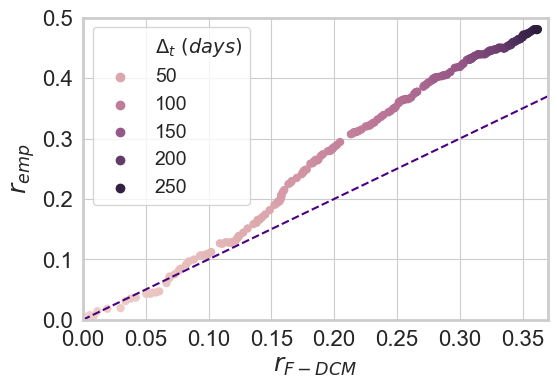}
         \caption{2007}
         \label{Fig:rec_2007}
     \end{subfigure}
\caption{Expected reciprocity by F-DCM $r_{F\text{-}DCM}$ vs. empirical reciprocity $r_{emp}$ w.r.t. different aggregation periods, in 1999 and 2007.}
\label{Fig:rho_cimi_detailed}
\end{figure}

\begin{figure}[H]
     \centering
     \begin{subfigure}[h!]{0.7\textwidth}
         \centering
         \includegraphics[width=\textwidth]{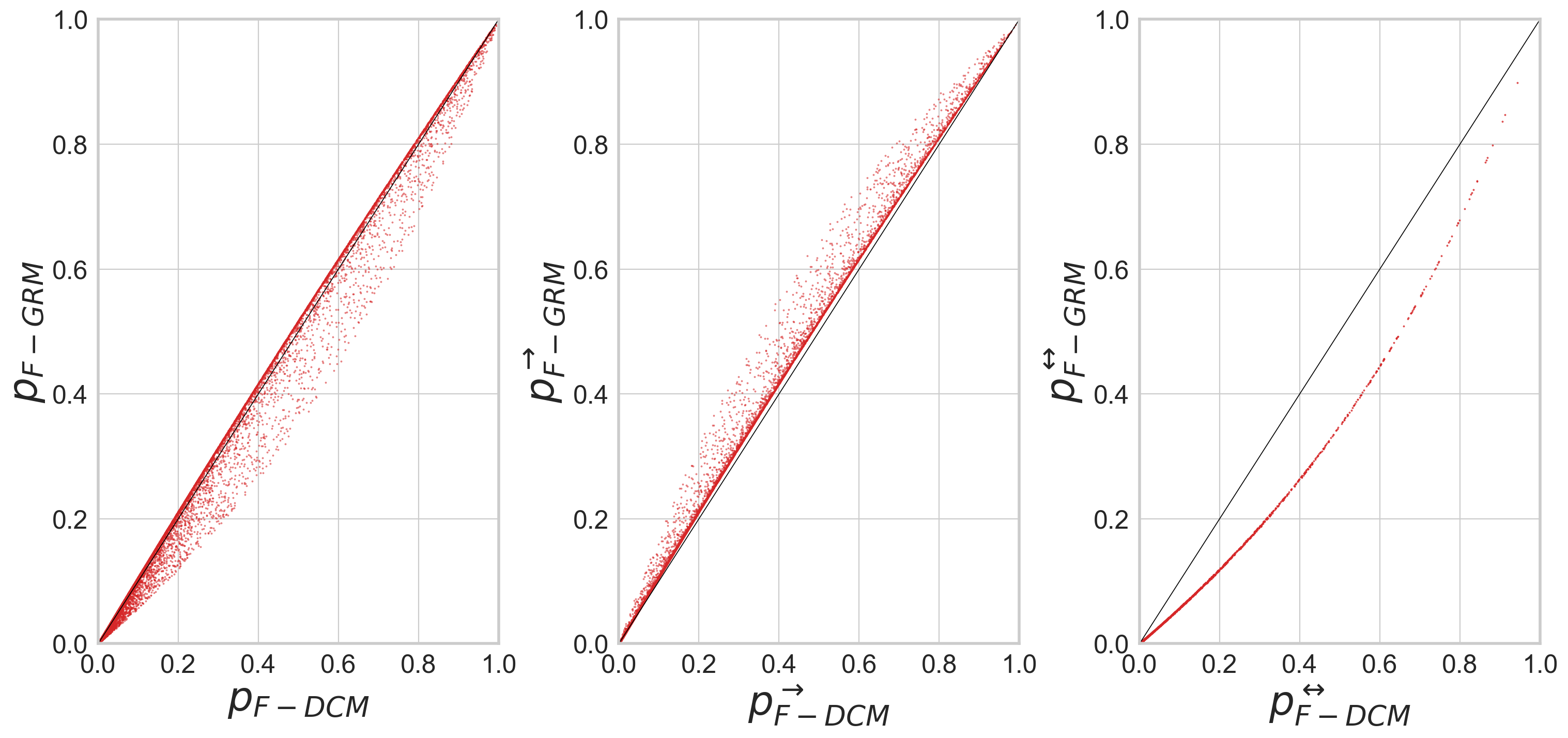}
         \caption{$\Delta_t=17$ days, $\rho_{F\text{-}DCM}=-0.05$}
         \label{Fig:pij-a}
     \end{subfigure}
     \begin{subfigure}[h!]{0.7\textwidth}
         \centering
         \includegraphics[width=\textwidth]{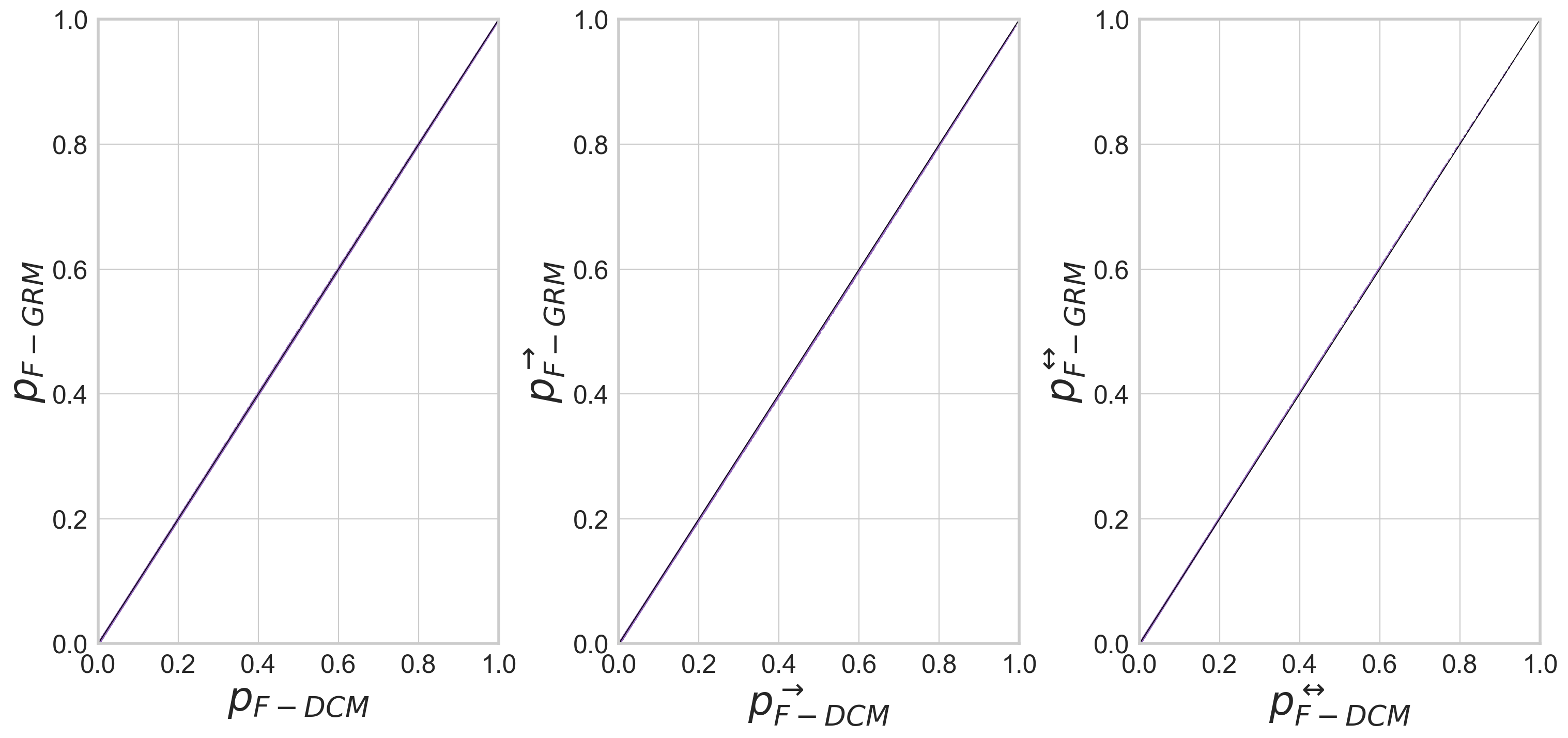}
         \caption{$\Delta_t=45$ days, $\rho_{F\text{-}DCM}=0.002$}
         \label{Fig:pij-b}
     \end{subfigure}
     \begin{subfigure}[h!]{0.75\textwidth}
         \centering
         \includegraphics[width=\textwidth]{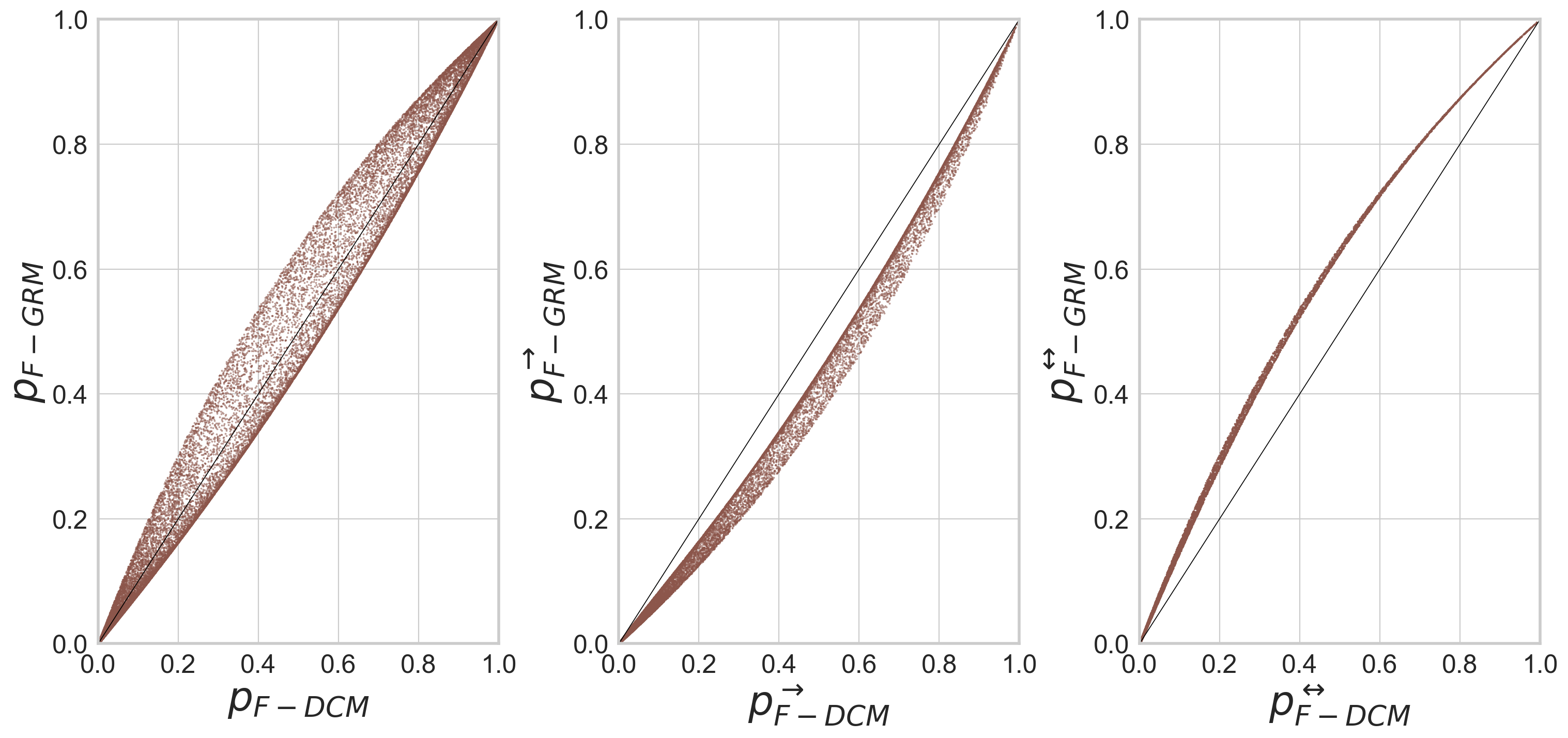}
         \caption{$\Delta_t=256$ days, $\rho_{F\text{-}DCM}=0.13$}
         \label{Fig:pij-c}
     \end{subfigure}     
        \caption{Link probability, 1999. Comparison between F-DCM and F-GRM. On the left there are the unconditional probabilities, on the center there are the mono-directed probabilities and on the right, there are the bi-directed probabilities. Different colors correspond to different aggregation periods $\Delta_t$: in red (top) and brown (bottom) there are the cases of the minimum and maximum value of $\rho_{F\text{-}DCM}$ while in purple (center) there is the case of $\rho_{F\text{-}DCM}\sim0$.}
        \label{Fig:pij_1999}
\end{figure}

Fig. \ref{Fig:rho_cimi_detailed} provides a more detailed illustration of the relationship between the expected reciprocity by F-DCM ($r_{F\text{-}DCM}$) and the empirical reciprocity ($r_{emp}$) as the aggregation period varies. Notably, we observe distinct patterns in different years. In 1999, F-DCM over (under) estimates the empirical reciprocity for short (long) aggregation periods, while matching it for quarterly networks. Conversely, in 2007, F-DCM accurately reproduces the empirical reciprocity up to quarterly networks, after which the model begins to underestimate it. For the sake of readability, we present the results for 1999 and occasionally for 2007 in this section. Additional plots can be found in Appendix \ref{appendix:plots}. 
For a more systematic analysis of F-DCM's performance, Table \ref{tab:cross_times} presents the aggregation periods corresponding to the minimum, maximum, and values closest to zero for $\rho_{F\text{-}DCM}$ in each year. $t_{min}$ and $t_0$ undergo significant changes across the years. They exhibit relative stability during the period 1999-2006, undergo variations in the pre-crisis and crisis periods of 2007-2008, and further adjustments in 2012 after the LTRO measures.
Viewed from a different angle, this table reinforces the earlier observation. The value of the link reciprocity and its dynamical evolution are difficult to match and guess when only the link density is given. Indeed, $t_0$ varies from a monthly to a quarterly range. This further underscores the necessity of introducing an additional parameter related to reciprocity, as presented in our model, F-GRM, to complement F-DCM.
\begin{table}[h]
\resizebox{\textwidth}{!}{
    \centering
    \begin{tabular}{c|cccccc|c|ccccc}
\toprule
 year & $\rho_{F\text{-}DCM}^{min}$ & $t_{min}$ & $t_0$ & $t_{max}$&$\rho_{F\text{-}DCM}^{max}$&&year&$\rho_{F\text{-}DCM}^{min}$& $t_{min}$ & $t_0$ & $t_{max}$&$\rho_{F\text{-}DCM}^{max}$ \\
\midrule
1999 & -0.05 & 17 & 45 & 256 & 0.13 && 2007 & -0.01 & 15 & 17 & 245 & 0.19 \\
2000 & -0.04 & 14 & 52 & 253 & 0.16 && 2008& -0.01 & 8 & 18 & 255 & 0.18 \\
2001 & -0.02 & 17 & 47 & 250 & 0.16 && 2009 &-0.01 & 8 & 13 & 242 & 0.10 \\
2002 & -0.06 & 7 & 59 & 254 & 0.14 &&  2010 & -0.03 & 17 & 27 & 257 & 0.14 \\
2003 & -0.04& 16 & 52 & 253 & 0.20 &&  2011 & -0.02 & 13 & 37 & 238 & 0.11\\
2004 & -0.04 & 24 & 47 & 258 & 0.18&&  2012 & -0.007 & 9 & 16 & 243 & 0.14\\
2005 & -0.03 & 11 & 22 & 252 & 0.22&&  2013 & -0.02 & 28 & 52 & 248 & 0.11 \\
2006 & -0.02 & 10 & 37 & 254 & 0.26&&  2014 & -0.015 & 11 & 20 & 255 & 0.13\\
\bottomrule
\end{tabular}
}
    \caption{Aggregation periods that correspond to the minimum ($t_{min}$), closest to zero ($t_0$) and maximum ($t_{max}$) value of $\rho_{F\text{-}DCM}$ in each year in the period 1999-2014. In a year, if there are multiple crossing times ($t_0$) due to fluctuations of $\rho_{F\text{-}DCM}$ around the zero (see Fig. \ref{Fig:rho_cimi}), $t_0$ is the maximum one. }
    \label{tab:cross_times}
    
\end{table}
In Figs. \ref{Fig:pij_1999}, we present a comparison of link probabilities between F-DCM and our model F-GRM, considering three different aggregation periods. Figs. \ref{Fig:pij-a} and \ref{Fig:pij-c} showcase the cases of the minimum and maximum values of $\rho_{F\text{-}DCM}$, while Fig. \ref{Fig:pij-b} represents the case where $\rho_{F\text{-}DCM}\sim0$. On the left, we illustrate the F-DCM unconditional probability $p_{ij}^{F\text{-}DCM}$ (Eq. \ref{eq:pij_cimi}) versus the F-GRM probability $p_{ij}^{F\text{-}GRM}$ (Eq. \ref{eq:p_grm}). In the center, we show the mono-directed probability $p_{ij}^{ F\text{-}DCM\rightarrow}=p_{ij}^{F\text{-}DCM}~(1-p_{ji}^{ F\text{-}DCM})$ with respect to $p^{F\text{-}GRM\rightarrow}$, and on the right, we illustrate the bi-directed probability $p_{ij}^{F\text{-}DCM\leftrightarrow}=p_{ij}^{F\text{-}DCM}p_{ji}^{F\text{-}DCM}$ versus $p_{ij}^{F\text{-}GRM\leftrightarrow}$.
Since both models constrain the total number of links $L=\sum_{i\neq j} p_{ij}$, the sum of the unconditional link probability is the same in both models, even if they are differently distributed. What changes is the repartition into mono and bidirected probabilities. As expected, when $\rho_{F\text{-}DCM}\sim0$, in the center, the link probabilities of the two models overlap, demonstrating that our extension aligns with the state-of-the-art. In the cases of negative and positive $\rho_{F\text{-}DCM}$, we observe opposite distributions of link probabilities into mono and bidirected probabilities.

Using link probabilities predicted by the two models, we compare in Appendix \ref{Appendix:accuracy} the accuracy of F-DCM and F-GRM in reconstructing single links using both ROC curve and cross-entropy loss as evaluation metrics. We find that both models have comparable performances on average, with F-GRM systematically outperforming F-DCM at large aggregation scales. Since F-GRM uses the information on global reciprocity for reconstruction, it is expected that it outperforms the F-DCM benchmark when employing higher-order network metrics rather than ROC curve and cross-entropy loss, which consider single links only for comparison. For this reason, we present below a comparative analysis between the two models in terms of spectral properties.

\subsection{Spectral Properties}
\begin{figure}[h]
     \centering
     \begin{subfigure}[b]{0.49\textwidth}
         \centering
         \includegraphics[width=\textwidth]{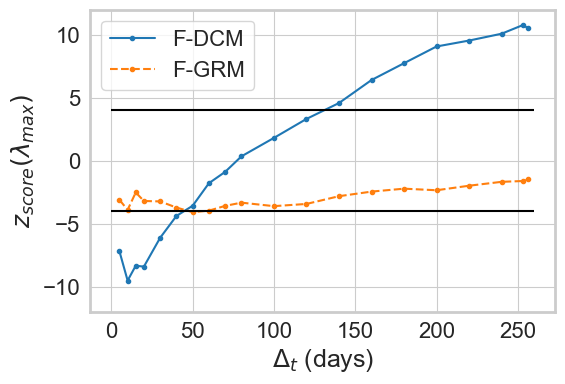}
         \caption{1999}
         \label{Fig:z_1999}
     \end{subfigure}
     %\hfill
     \begin{subfigure}[b]{0.49\textwidth}
         \centering
         \includegraphics[width=\textwidth]{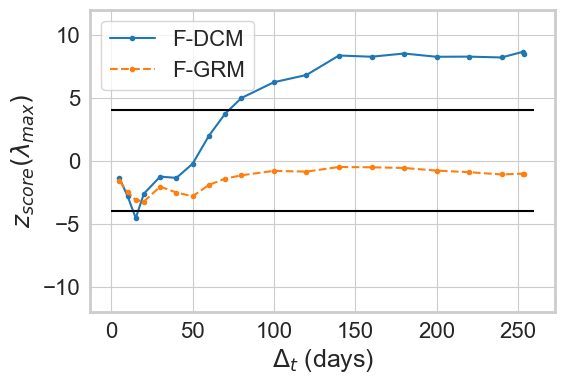}
         \caption{2007}
         \label{Fig:z_2007}
     \end{subfigure}
\caption{$z_{score}$ of the maximum eigenvalue as the aggregation period increases, 1999 and 2007. Horizontal black lines show the values $z_{score}\pm 4$. F-DCM in blue and F-GRM in orange.}
\label{Fig:z_scores}
\end{figure}

The interbank market model is depicted as a network, providing a visual representation of lending interactions among its constituents and facilitating an understanding of how its stability is influenced by the underlying topology. To mitigate risk and instability in a dynamic system, it is crucial to identify the topological properties that stabilize the interaction network. Graph spectra, particularly the leading eigenvalue $\lambda_{max}$, play a key role in terms of systemic stability, as they account for the loops and cycle structure determining the propagation and amplification of an initial shock.
However, assessing the systemic risk of the interbank network poses an additional challenge. Since we lack direct access to the empirical network, we must reconstruct it and ensure that the crucial graph spectral properties of the empirical network are faithfully preserved. This guarantees that the ensemble of reconstructed networks serves as a reliable proxy for the empirical one, particularly concerning stability properties.
The results presented below demonstrate that our model outperforms the state-of-the-art (F-DCM) in terms of spectral properties of the adjacency matrix, namely the distribution of the eigenvalues of $A$ at different aggregations. The comparison involves generating an ensemble of 1000 realizations for both F-DCM and F-GRM.
\paragraph{Maximum Eigenvalue} To assess the performance of F-DCM and F-GRM, Fig. \ref{Fig:z_scores} presents the $z_{score}$ of the leading eigenvalue, defined as:
\begin{equation}
    z_{score}(\lambda_{max})=\frac{\lambda_{max}^{emp}-mean(\overline{\lambda}_{max}^{~ens})}{std(\overline{\lambda}_{max}^{~ens})}
    \label{eq:z-score}
\end{equation}
where $\lambda_{max}^{emp}$ is the maximum eigenvalue of the empirical adjacency matrix, and $\overline{\lambda}_{max}^{~ens}$ is the list of the maximum eigenvalues of each adjacency matrix in the generated ensemble.
We successfully validate the normal distribution of the $z_{score}(\lambda_{max})$ through the Kolmogorov-Smirnov test.
In the case of F-GRM, the $z_{score}$ remains almost constant and negative as the aggregation period increases. This indicates that this model consistently overestimates the empirical maximum eigenvalue. Notably, F-GRM performs better in 2007 than in 1999, and its absolute value remains below 4 even in 1999.
For F-DCM, a similar trend to $\rho_{F\text{-}DCM}$ is observed. As the aggregation period increases, the $z_{score}(\lambda_{max})$ reaches the minimum, then zero, and finally the maximum at the yearly level. When considering $|z_{score}(\lambda_{max})|<4$ as the confidence interval, F-DCM fails to reconstruct the empirical maximum eigenvalue when the aggregation period is below quarterly or above biannual. Conversely, in the case of F-GRM, the value of $z_{score}(\lambda_{max})$ always falls within the confidence interval. Additionally, the maximum eigenvalue is consistently overestimated, ensuring that the systemic risk associated with the generated network is never underestimated compared to the empirical networks.
The observations in Fig. \ref{Fig:z_scores} are consistent across other yearly plots in the period 1999-2014 that are reported in Appendix \ref{appendix:plots}. To assess systemic risk effectively, it is important to have a model with stable performance in terms of under/overestimation. In this context, F-GRM emerges as a robust reconstruction model,
since it provides an upper bound in terms of spectral properties w.r.t. the empirical ones and it always falls within the confidence interval.
It is important to note that F-DCM performs similarly to F-GRM in aggregation periods where $\rho_{F\text{-}DCM}\sim 0$. However, since these periods vary over the years, providing a general recommendation on whether F-DCM or F-GRM is more suitable becomes challenging.
Despite F-GRM outperforming F-DCM, a practical approach might be to conserve computing resources by utilizing F-DCM when the aggregation period is approximately quarterly (i.e., when $\rho_{F\text{-}DCM}\sim 0$), while employing F-GRM for lower and higher frequency periods.
\paragraph{Bulk of the spectrum} 
In the case of the interbank networks, nodes are not equivalent both in terms of interbank assets and liabilities and also the link probability in F-DCM and F-GRM takes into account these heterogeneities. 
Regarding the properties of the bulk of the spectrum, there are theoretical results regarding the random directed matrices. In \cite{sommers1988spectrum}, the authors examine an ensemble of $N\times N$ large random real asymmetric matrices $J$. These matrices are defined by a Gaussian distribution with a zero mean and correlations: $N[J_{ij}]J=1$ and $N[J_{ij}J_{ji}]_J=\tau$, where the brackets $[\dots]_J$ denote the ensemble average, and $-1\leq \tau\leq 1$. The study reveals that the average eigenvalue distribution is uniform in an ellipse in the complex plane, with the real and imaginary axes being $1+\tau$ and $1-\tau$, respectively. The correlation $\tau$ is also associated with link reciprocity $r$: in the case of an antisymmetric network, $\tau=-1$ and $r=0$, while $\tau=1$ and $r=1$.

We want now to adapt these results in our heterogeneous case of the interbank network. To meet the initial assumption, we first need to rescale the adjacency matrix $A$ to fulfil the initial conditions of mean and correlations of the matrix ensemble in \cite{sommers1988spectrum}. Details on the calculations can be found in Appendix \ref{appendix:ellipse}. The resulting expression is: 
\begin{equation}
    \widetilde{J}_{ij}=\frac{A_{ij}-p_{ij}}{\sqrt{Np_{ij}(1-p_{ij})}}
\end{equation}
\begin{equation}    \braket{\widetilde{J}_{ij}\widetilde{J}_{ji}}=\frac{p^\leftrightarrow_{ij}-p_{ij}p_{ji}}{N\sqrt{p_{ij}(1-p_{ij})p_{ji}(1-p_{ji})}}=\frac{\tau_{ij}}{N}
\label{eq:tauij}
\end{equation}
where $p_{ij}$ is the unconditional link probability, and $p^{\leftrightarrow}_{ij}$ is the bidirected link probability of the considered method. 
In \cite{sommers1988spectrum}, the correlation $\tau$ is not node-dependent, unlike the interbank network case, as shown in Eq. \ref{eq:tauij}. In the case of F-DCM, each link is independently sampled, so $p_{ji}^{\leftrightarrow F\text{-}DCM}=p_{ij}^{F\text{-}DCM}p_{ji}^{F\text{-}DCM}$ and $\tau_{ij}=0~\forall i,j$.

\begin{figure}[ht]
     \centering
     \begin{subfigure}[h!]{0.3\textwidth}
         \centering
         \includegraphics[width=\textwidth]{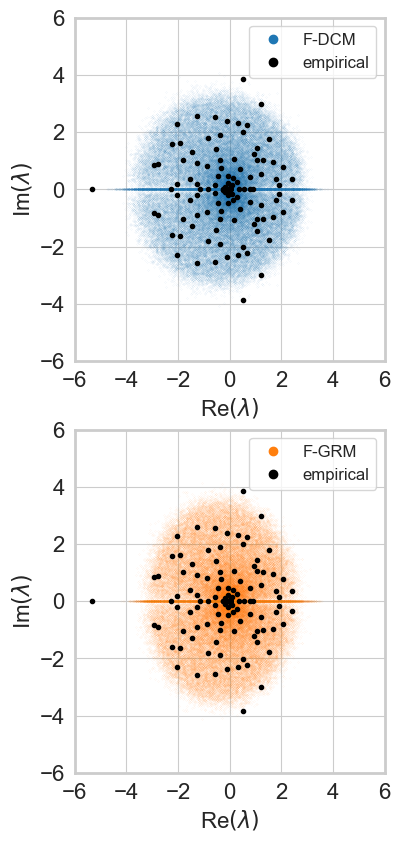}
         \caption{$\Delta_t=17$ days,\\$\rho_{F\text{-}DCM}=-0.05$}
         \label{Fig:ellipse-a}
     \end{subfigure}
     \begin{subfigure}[h!]{0.3\textwidth}
         \centering
         \includegraphics[width=\textwidth]{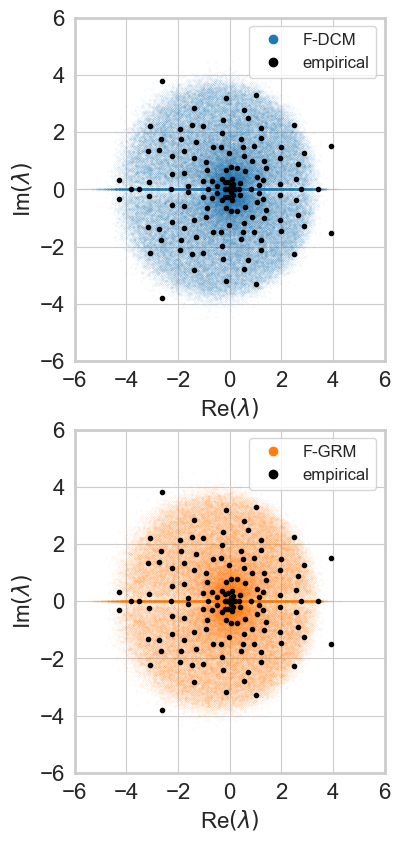}
         \caption{$\Delta_t=45$ days,\\$\rho_{F\text{-}DCM}=0.002$}
         \label{Fig:ellipse-b}
     \end{subfigure}
     \begin{subfigure}[h!]{0.3\textwidth}
         \centering
         \includegraphics[width=\textwidth]{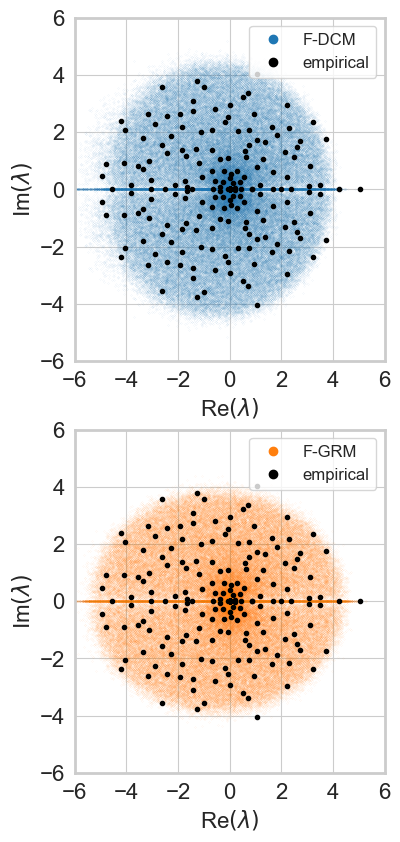}
         \caption{$\Delta_t=256$ days,\\$\rho_{F\text{-}DCM}=-0.13$}
         \label{Fig:ellipse-c}
     \end{subfigure}     
        \caption{Bulk of the spectra of the adjacency matrix in the complex plane, 1999. Different colors correspond to different models: F-DCM in blue and F-GRM in orange. Black dots represent the eigenvalue of the empirical adjacency matrix. On the left and the right, there are the cases of the minimum and maximum value of $\rho_{F\text{-}DCM}$, respectively, while in the center there is the case of $\rho_{F\text{-}DCM}\sim0$.}
        \label{Fig:ellipseaij}
\end{figure}

What F-GRM imposes is the global reciprocity $r$ but this also has an effect at the node level.  $\tau_{ij}$ is instead a characteristic of the model, that reflects how the global reciprocity is distributed over pairs of nodes and its value depends on the nodes' fitness, i.e., their heterogeneity in total interbank assets and liabilities; for more details, see the Appendix \ref{appendix:ellipse}. In Fig. \ref{Fig:hist_tau}, we present the distribution of the correlation $\tau_{ij}$; different colors correspond to different aggregation periods in 1999.
\begin{figure}[ht]
    \centering
    \includegraphics[width=0.75\textwidth]{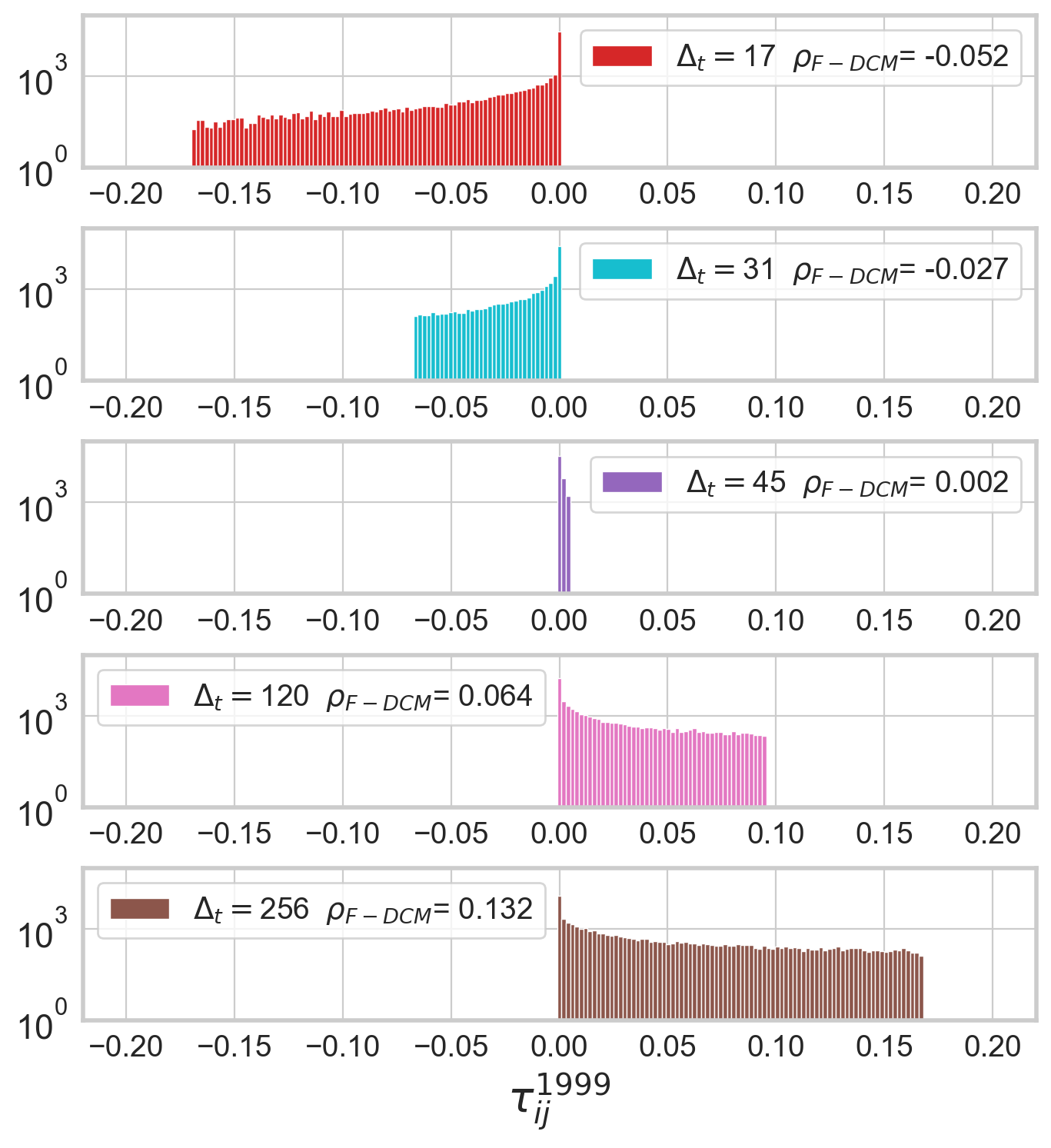}
    \caption{Histogram of $\tau_{ij}$, 1999. Different colors correspond to different aggregation periods $\Delta_t$ (days): in red and brown there are the cases of the minimum and maximum value of $\rho_{F\text{-}DCM}$ while in purple there is the case of $\rho_{F\text{-}DCM}\sim0$.}
    \label{Fig:hist_tau}
\end{figure}
A correlation is observed between the $\tau_{ij}$ distribution and the value of $\rho_{F\text{-}DCM}$; not only all $\tau_{ij}$ values have the same sign as $\rho_{F\text{-}DCM}$, but the amplitude of the $\tau_{ij}$ distribution increases as the absolute value of $\rho_{F\text{-}DCM}$ increases. 
For completeness, Fig. \ref{Fig:ellipseaij} reports the bulk of the spectra of the ensemble of the generated adjacency matrices; F-DCM in blue and F-GRM in orange. Black dots represent the eigenvalues of the empirical network. The aggregation period that corresponds to the minimum (maximum) value of $\rho_{F\text{-}DCM}$ is shown in Fig. \ref{Fig:ellipse-a} (Fig. \ref{Fig:ellipse-c}). Fig. \ref{Fig:ellipse-b} reports the case of $\rho_{F\text{-}DCM}\sim 0$. A comparison with the theoretical results in \cite{sommers1988spectrum} can be made, taking into account that Fig. \ref{Fig:ellipseaij} reports the bulk of the adjacency matrix $A$, not the rescaled matrix $\tilde{J}$. In the case of F-DCM, we observe that the bulk of the spectra forms a circle for any aggregation period, consistent with the results in \cite{sommers1988spectrum}, where $\tau_{ij}$ is equal to zero, and the real and imaginary axes are equal. Similarly, in the case of F-GRM, the results align with those in \cite{sommers1988spectrum}. At the bottom of Fig. \ref{Fig:ellipse-a} (Fig. \ref{Fig:ellipse-c}), we have the case of negative (positive) $\rho_{F\text{-}DCM}$ and $\tau_{ij}$ (see red (brown) histogram in Fig. \ref{Fig:hist_tau}), corresponding to an ellipse whose imaginary axis is longer (shorter) than the real one. Analogously, at the bottom of Fig. \ref{Fig:ellipse-b}, we recover the circular shape. Examining Fig. \ref{Fig:ellipseaij}, we also observe that the elliptical bulk of the spectrum in the F-GRM case better fits the eigenvalues of the empirical network.

\clearpage
\section{Conclusions}
In this paper, we present an extension of the state-of-the-art F-DCM, which incorporates constraints not only on link density but also on global link reciprocity. Although our case study focuses on the interbank market network, our proposed model, F-GRM, is versatile and applicable in any network context characterized by data scarcity.

Our model contributes to the literature on reconstruction methods in several ways. Firstly, we observe that reciprocity strongly varies with different aggregation periods, and it cannot be inferred from link density by analyzing empirical interbank market networks (e-MID data). This empirical evidence supports our objective of enhancing F-DCM by constraining not only the link density but also the reciprocity, as the latter provides additional information to the model compared to the former. Secondly, we demonstrate that F-GRM generates an ensemble of networks with spectral properties closer to empirical ones. This is crucial because the spectra of graphs, particularly the leading eigenvalue $\lambda_{max}$, play a key role in terms of systemic stability by accounting for loops and cycle structures that influence the propagation and amplification of an initial shock. Thirdly, when assessing system stability, it is important to have a model with stable performance in terms of under/overestimation of the maximum eigenvalue and thus systemic risk. In this regard, F-GRM is a reliable reconstruction model, generating an ensemble of networks with spectral properties consistently upper-bounding empirical ones and always within the confidence interval. We show that our model effectively preserves spectral properties, showing comparable performance to F-DCM when the aggregation period is approximately quarterly, i.e. when the expected global reciprocity is close to the empirical one. In contrast, F-GRM outperforms F-DCM when the empirical reciprocity deviates from F-DCM expectations, occurring for periods shorter or longer than quarterly.

In conclusion, we propose an extension of F-DCM that also considers link reciprocity. We demonstrate that our model, F-GRM, outperforms F-DCM, especially in resembling the spectral properties of the empirical adjacency matrices. It is worth noting that our model can be applied to directed networks of any type and can be particularly useful when the empirical network exhibits significant under- or over-expression of reciprocity compared to the hypothesis of random connections.

Future work involves integrating reciprocity into existing reconstruction models for weighted matrices. In the case of the interbank network, the only local information available is the total interbank assets and liabilities, representing nodes' out- and in-strength. To reconstruct bilateral exposure from these aggregated measures, additional reconstruction models must be considered. These models can be deterministic, such as the IRF and RAS algorithms \cite{upper2004estimating,macchiati2022systemic}, or probabilistic, as demonstrated in \cite{cimini2015systemic,parisi2020faster}. 
All of these models require the adjacency matrix as input, so F-GRM could be considered an enhanced plug-and-play attachment that incorporates link reciprocity when imputing the weighted structure of these models. A potential future research direction entails expanding the current reconstruction method for weighted matrices to include weighted reciprocity, discerning between mono- and bi-directed weighted links. This extension follows a similar approach to that of F-GRM for adjacency matrices.

\section*{Acknowledgements}
We thank G. Cimini, F. Lillo and T. Squartini for the useful discussions. The authors thank Scuola Normale Superiore, Pisa, for providing the e-MID data. VM and DG acknowledge support from the project NetRes - `Network analysis of economic and financial resilience', Italian DM n. 289, 25-03-2021 (PRO3 Scuole), CUP D67G22000130001 (\url{https://netres.imtlucca.it}) by the Scuola Normale Superiore in Pisa, the IMT School of Advanced Studies Lucca, and the Sant’Anna School of Advanced Studies in Pisa.
This work is also supported by the European Union - NextGenerationEU - National Recovery and Resilience Plan (Piano Nazionale di Ripresa e Resilienza, PNRR), project `SoBigData.it - Strengthening the Italian RI for Social Mining and Big Data Analytics' - Grant IR0000013 (n. 3264, 28/12/2021) (\url{https://pnrr.sobigdata.it/}). PM acknowledges financial support under the National Recovery and Resilience Plan (NRRP), Mission 4, Component 2, Investment 1.1, Call for tender No. 104 published on 2.2.2022 by the Italian Ministry of University and Research (MUR), funded by the European Union – NextGenerationEU– Project Title “Realized Random Graphs: A New Econometric Methodology for the Inference of Dynamic Networks” – CUP B53D23010100001 - Grant Assignment Decree No. 2022MRSYB7\_02 by the Italian Ministry of Ministry of University and Research (MUR).
\appendix
\clearpage
\section{Parameters' estimation procedure}\label{appendix:trf}
Exponential Random Graph Models (ERGMs) are usually estimated using network data by using the generalized method of moments \cite{hansen1982large}: the parameters of the models are estimated by matching the observed values of network metrics with their expected values over the network ensemble. In practice, the estimation process consists of solving a system of nonlinear equations defined in a constrained parameter space. The parameter domain\footnote{Parameters, being the exponential of Lagrange multipliers, are constrained in the positive real domain.} is constrained to ensure that the generic link probability remains positive and within the range of 0 to 1.
The solution of the nonlinear equations is found using the Trust Region Reflective (TRF) Algorithm with Bound Constraints\footnote{TRF algorithm is implemented in the python library \textit{scipy} \cite{2020SciPy-NMeth}, in the least squares method \textit{scipy.optimize.least\_squares}.} whose pseudocode is available in Alg. \ref{ALG}.
\begin{algorithm}[h]
\caption{~\\ Trust Region Reflective (TRF) Algorithm  with Bound Constraints}
\SetAlgoLined
\KwIn{$f$, $x_0$=\textbf{$\mathbb{1}$}, bounds=(0,$\infty$)}
\KwOut{x} 
\BlankLine
$k \gets 0$\;
\While{not converged and $k <$ max\_iterations}{
  $\nabla f \gets \text{computeGradient}(\textbf{x})$\;
  $\textbf{H} \gets \text{computeHessian}(\textbf{x})$\;
  $\textbf{p} \gets \text{solveSubproblem}(\textbf{H}, \nabla f, \text{bounds})$\;
  $\alpha \gets \text{lineSearch}(\textbf{x}, \textbf{p})$\;
  $\textbf{x} \gets \textbf{x} + \alpha \textbf{p}$\;
  converged $\gets \text{checkConvergence}(\nabla f, \alpha, \textbf{p})$\;
  $k \gets k + 1$\;
}
\label{ALG}
\end{algorithm}
~
\\\\It is worth remembering that F-DCM has one parameter to tune and one constraint (link density), while F-GRM has two parameters to tune and two constraints (link density and reciprocity). The estimation process thus depends on solving a system of two (one) nonlinear equations defined in a constrained parameter space for F-GRM (F-DCM) as shown in Eq. \ref{eq:sysFGRM} (Eq. \ref{eq:sysFDCM}). Other ERG models are even more complex; the estimation process in the DCM, GRM, and RCM requires solving a system of $2N$, $2N+1$, and $3N$ equations, respectively, where $N$ is the number of nodes in the network.
\\
Despite F-GRM requiring solving an additional non-linear equation compared to F-DCM, the estimation process results as efficient. In the case of 1999, when the number of nodes is the highest ($N=212$) in the e-MID dataset, parameter tuning for all aggregation periods (256 snapshots) lasts 4 seconds in F-DCM and 35 seconds in F-GRM\footnote{These results are obtained by running the optimization of the two models on the same laptop (locally, not on a distributed server), by choosing the same tolerance, and setting the initial value of each parameter to one in both cases.}. We consider both computing times acceptable\footnote{The optimization algorithm to solve F-GRM could be further optimized, but that goes beyond the scope of this paper.} since each set of parameters has to be tuned only once for each network snapshot.

\section{e-MID Data}\label{appendix:data}
We have access to two distinct e-MID datasets spanning the period from 1999 to 2014, with an overlapping duration of Sept/2010 to 2012. Consequently, our initial task involves reconciling these datasets. Initially, we focus on overnight loans, constituting the majority (95\%) in terms of both transactions and volumes during the period 2010-2014. Subsequently, we address the challenge of reconciling the anonymized identification codes for loan takers and givers across the two data sources.
To address this, we concentrate on the overlapping period and consider the volumes of transactions unique to each day, creating a mapping between the differing anonymization keys. It is noteworthy that the two datasets exhibit slight discrepancies during the overlap period. Consequently, we retain only those transactions reported in both datasets and exclusively those banks with clearly identified mappings\footnote{In cases where the mapping is not unique, we assign the label of the most probable one (more than 95\%). One bank possesses a spurious mapping, leading to its removal from the dataset.} .
Our decision to map the anonymization of the second dataset into the first, rather than vice versa, is influenced not only by the longer coverage period of the first dataset but also by the increased activity of more banks in the e-MID during that timeframe. Specifically, only a bank reported in the second dataset is absent in the first, and given its involvement in only one transaction over the entire period, we have excluded it.
The resulting dataset encompasses 99\% (98\%) of the transactions and 99\% (93\%) of the volumes from the first (second) dataset.

\section{Single-link reconstruction performance of F-DCM and F-GRM}\label{Appendix:accuracy}
\begin{figure}[h]
  \centering
  \begin{subfigure}{0.7\textwidth}
    \centering
    \includegraphics[width=\linewidth]{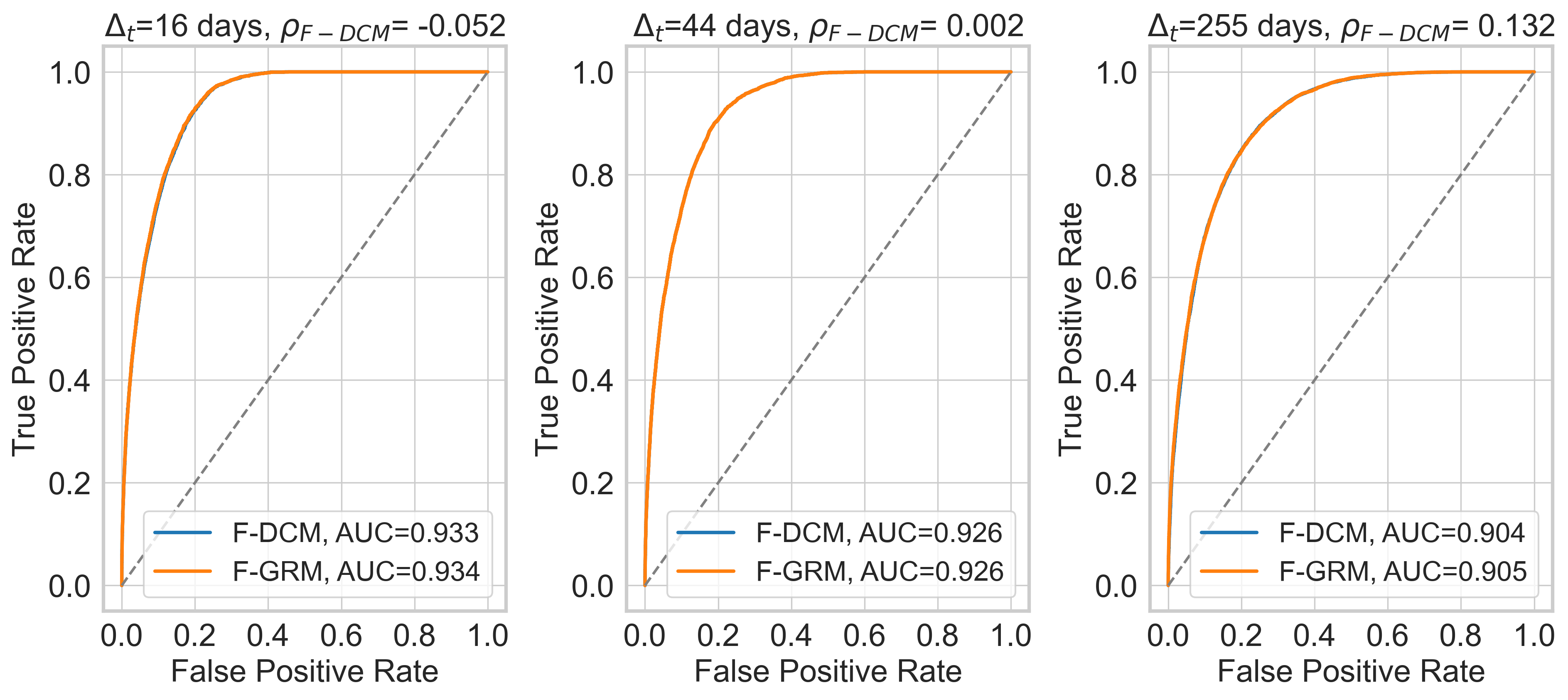} % replace 'figure1' with thefilename of your first figure
    \caption{1999} % Add your caption for the first figure here
    \label{fig:ROC_1999}
  \end{subfigure}
  \begin{subfigure}{0.7\textwidth}
    \centering
    \includegraphics[width=\linewidth]{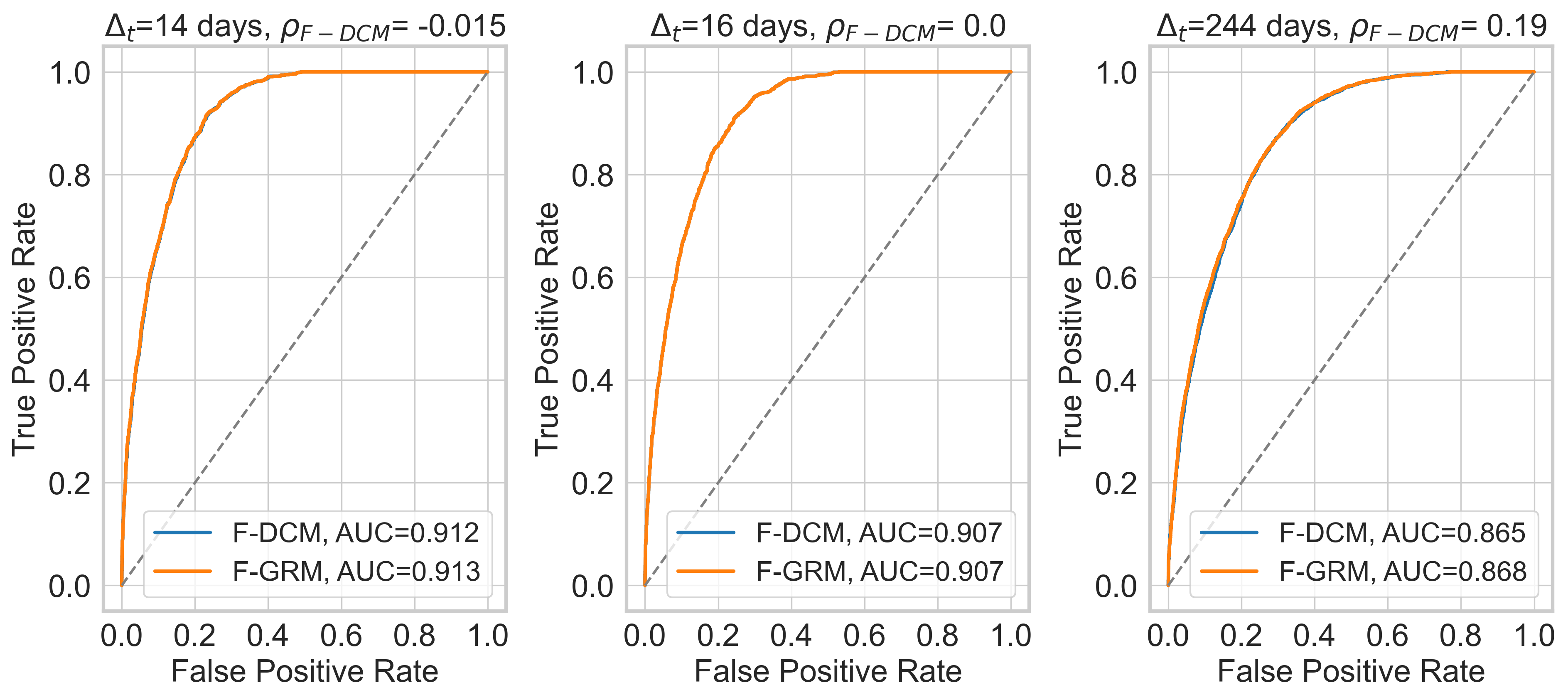} % replace 'figure2' with the filename of your second figure
    \caption{2007} % Add your caption for the second figure here
    \label{fig:Roc_2007}
  \end{subfigure}
  \caption{Receiver Operating Characteristic (ROC) Curve.}
  \label{Fig:ROC}
\end{figure}

\begin{figure}[h]
\centering
    \includegraphics[width=0.5\linewidth]{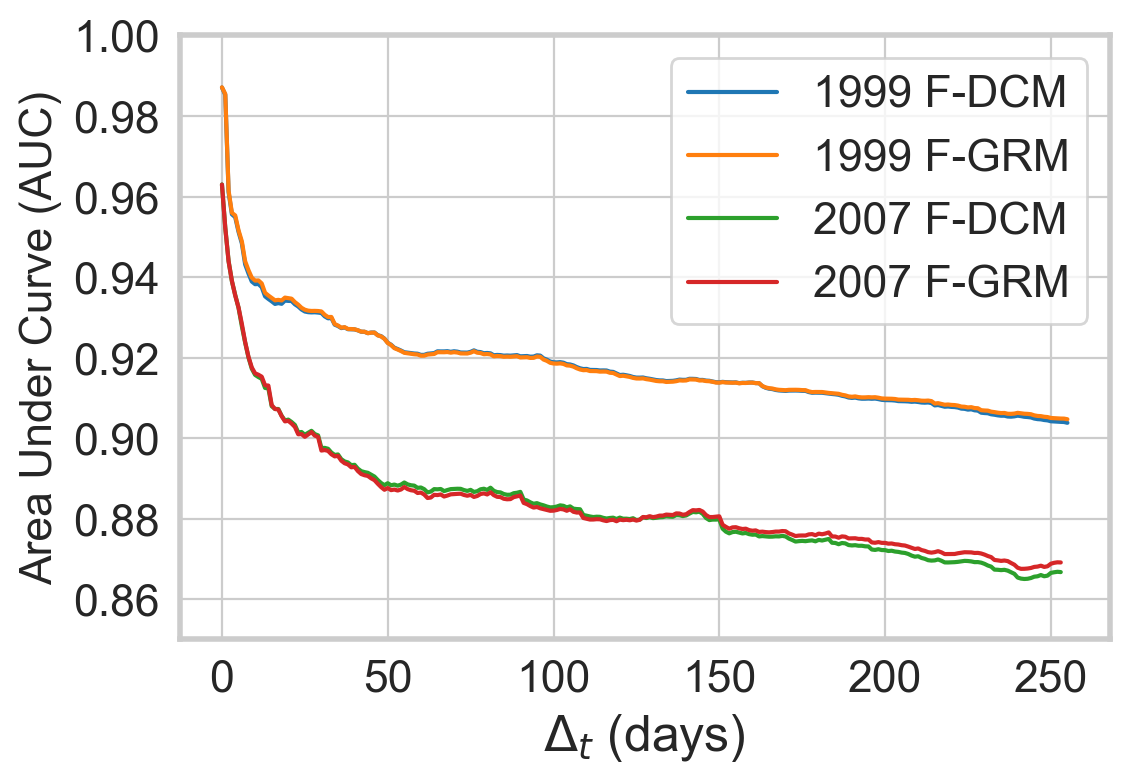} 
\caption{Comparison between Area Under Curve (AUC) in the case of F-DCM and F-GRM, 1997 and 2007.}
  \label{Fig:AUC}
\end{figure}

\begin{figure}[h]
\centering
    \includegraphics[width=0.7\linewidth]{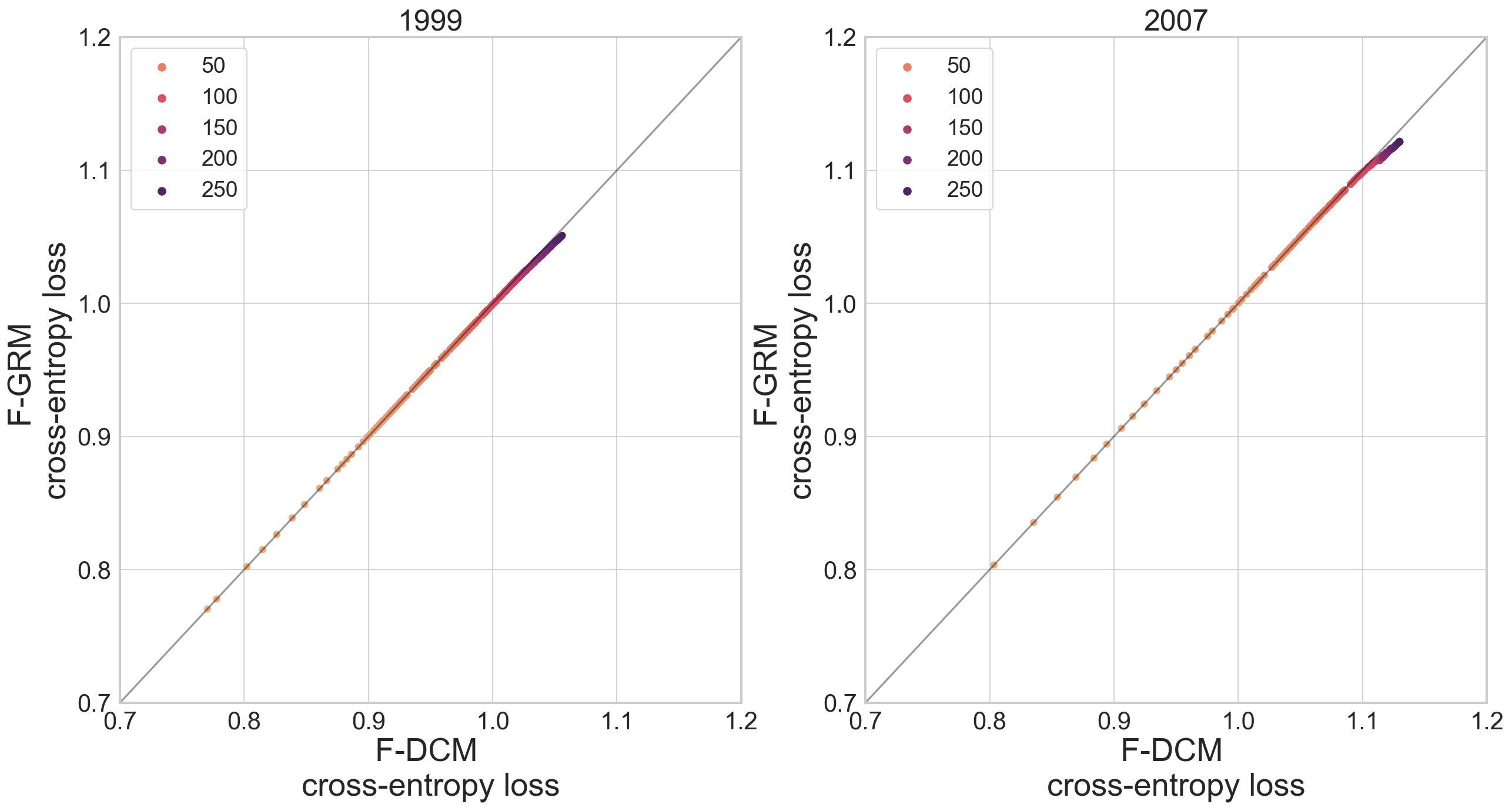} 
\caption{Cross Entropy loss in the case of F-DCM and F-GRM, 1997 and 2007. Different colors correspond to different aggregation periods (days).}
  \label{Fig:cross-entropy}
\end{figure}

We compare the link probabilities ($p_{ij}\in[0,1]$) as described by both F-DCM and F-GRM with the sequence of observed links ($A_{ij}\in\{0,1\}$) to analyze the accuracy of the network reconstruction in terms of ROC curves. We first compute the ROC curve for each aggregation period. We report in Fig. \ref{Fig:ROC}, the ROC Curve related to the three aggregation periods analyzed in the paper, i.e., the minimum, maximum, and closest to zero value of $r_{F\text{-}DCM}$, in 1999 and 2007. Notably, F-GRM slightly outperforms the F-DCM model in terms of AUC for negative and positive $\rho_{F-DCM}$ (and this result is systematic for large aggregation), see the left and right subplots of Fig. \ref{Fig:ROC}, while displaying comparable performance with F-DCM when $\rho_{F-DCM}=0$, see the middle plot of Fig. \ref{Fig:ROC}. The latter result is expected since both models predict (approximately) equal link probabilities. Moreover, in Figure \ref{Fig:AUC}, we analyze the Area Under Curve (AUC) values across different aggregation periods in 1999 and 2007. Again, F-GRM slightly outperforms F-DCM. Notice that the AUC for F-GRM is systematically larger than the one for F-DCM for large aggregation, signaling that the reconstruction is consistently better even if the signal is small.
We expand our analysis on the accuracy performance by considering a multiclass classification scenario with four classes. Given a couple of nodes $i,j$, we have four possible outcomes; mono-directed link from $i$ to $j$ or from $j$ to $i$, bi-directed link, no link. The cross-entropy loss evaluates the disparity between the link probabilities predicted by the models and the true labels provided by the empirical networks. In Fig. \ref{Fig:cross-entropy}, we find the reported values of the cross-entropy loss for both F-DCM and F-GRM. As a loss metric, lower values indicate better reconstruction accuracy. We observe that F-DCM and F-GRM demonstrate comparable performances in 1999 and 2007 across all aggregation periods. However, there is a slight performance improvement for F-GRM with longer aggregation periods.
\\\\
Since the F-GRM model aims to capture the presence of loops in network data (i.e., reciprocated links), while the F-DCM model does not, we expect that the former's superior performance would appear more clearly when using higher-order network metrics (as opposed to the ROC curve and cross-entropy loss, which considers single links only). For this reason, our focus in the paper shifts to metrics that take into account spectral and cycle structures. Moreover, these structures are pivotal for systemic stability, offering a more comprehensive approach that accounts for crucial factors beyond standard statistical comparisons. 

\clearpage
\section{Additional plots}\label{appendix:plots}
\subsection{Reciprocity}
\begin{figure}[h]
    \centering
    \includegraphics[width=0.88\textwidth]{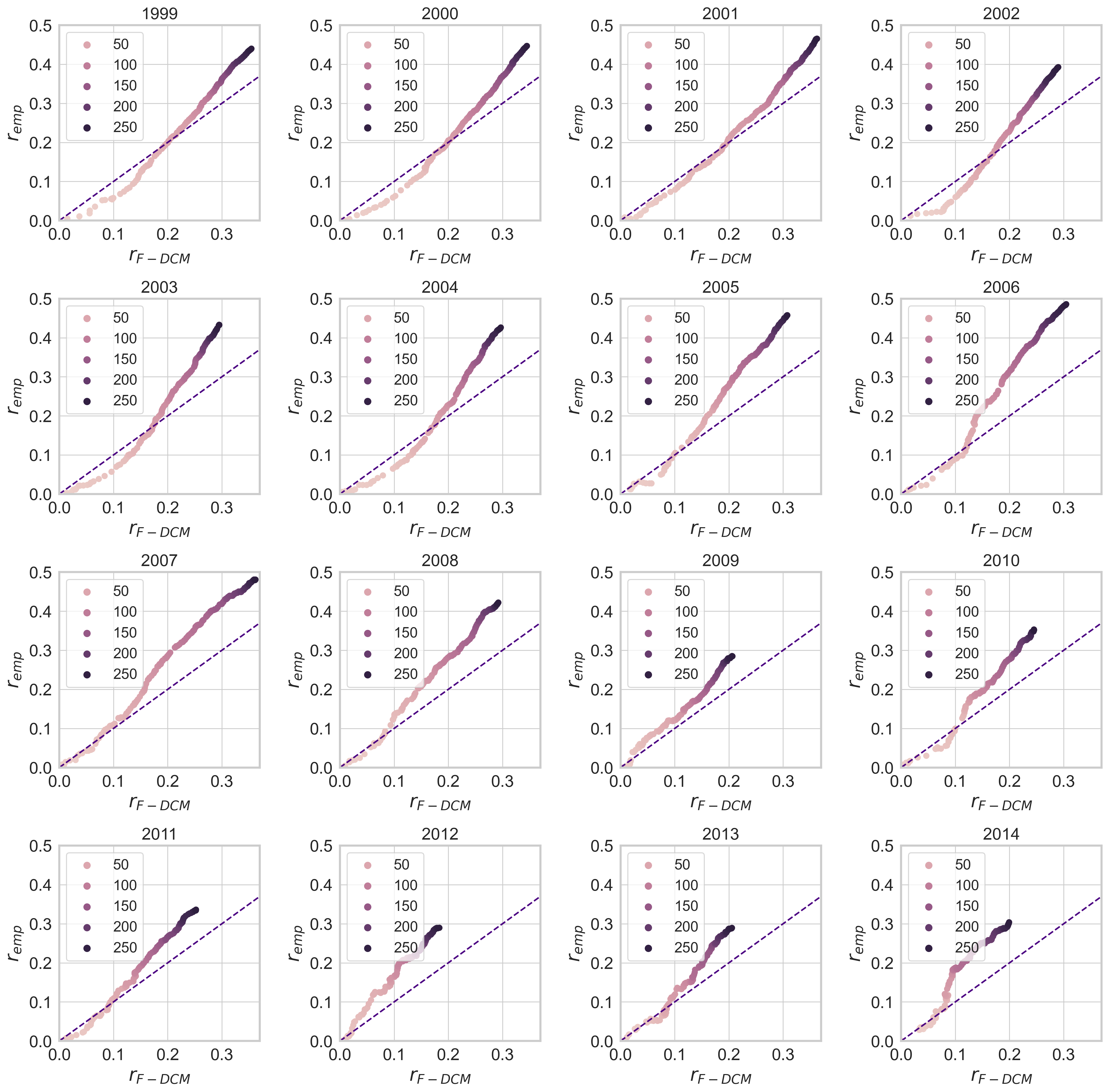}
    \caption{$r_{F\text{-}DCM}$ vs. empirical reciprocity $r_{emp}$ w.r.t. different aggregation periods in the years 1999-2014.}
    \label{r_appendix}
\end{figure}

Fig. \ref{r_appendix} shows the relationship between the expected reciprocity by F-DCM $r_{F\text{-}DCM}$ and the empirical one $r_{emp}$ as the aggregation periods vary, in the years 1999-2014.
\subsection{Link probability}
\begin{figure}[h]
     \centering
     \begin{subfigure}[h]{0.75\textwidth}
         \centering
         \includegraphics[width=\textwidth]{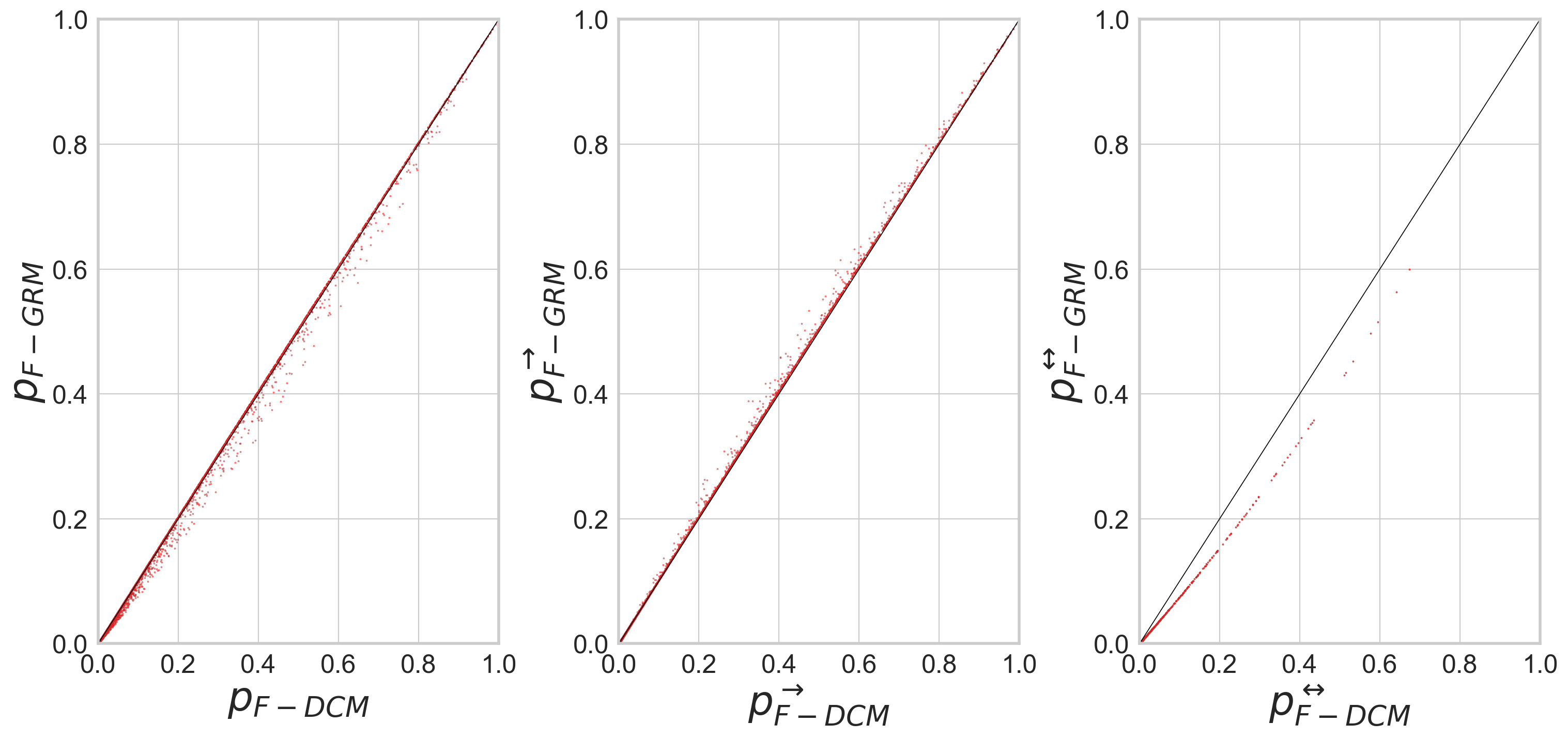}
         \caption{$\Delta_t=15$ days, $\rho_{F\text{-}DCM}=-0.015$}
         \label{Fig:pij2007-a}
     \end{subfigure}
     \begin{subfigure}[h]{0.75\textwidth}
         \centering
         \includegraphics[width=\textwidth]{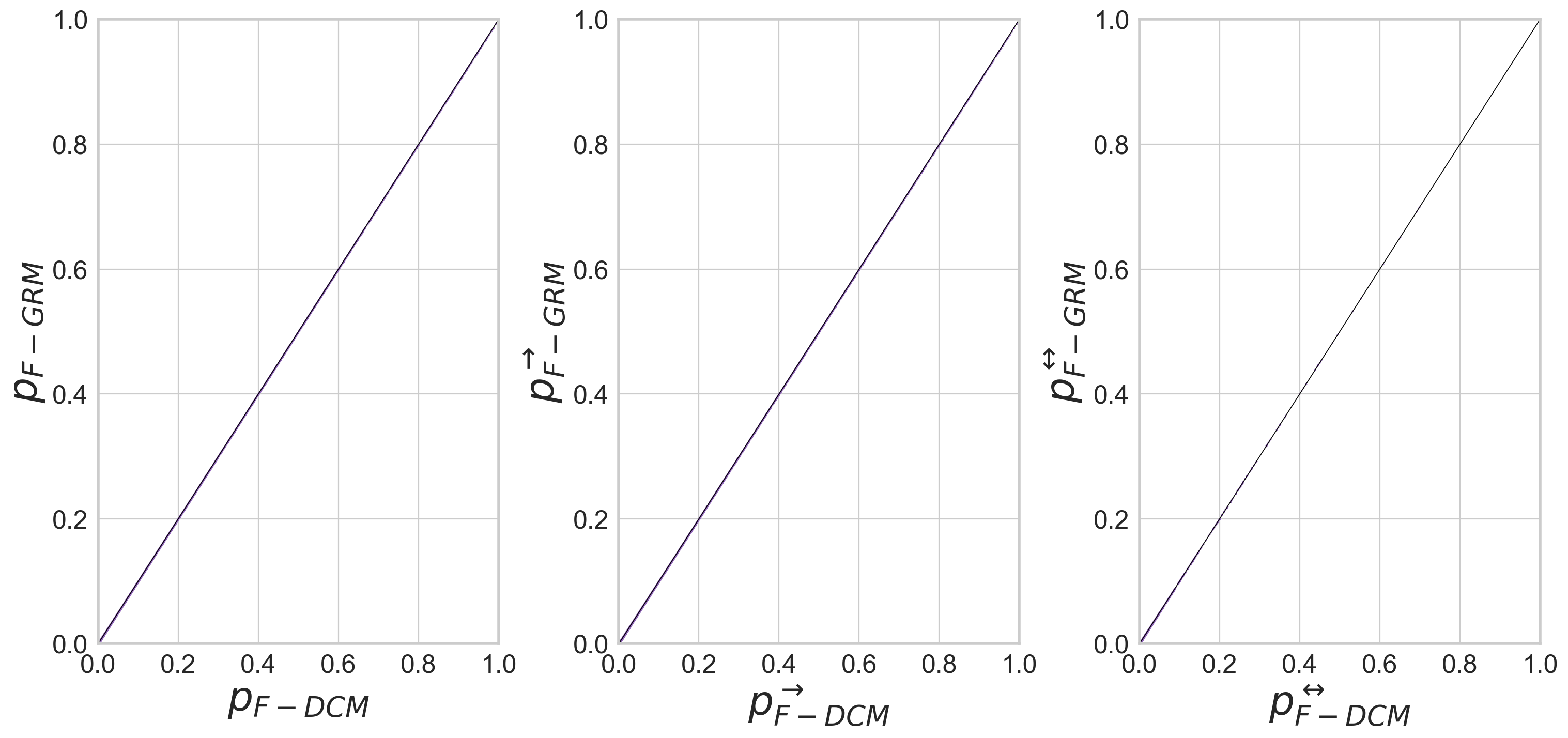}
         \caption{$\Delta_t=17$ days, $\rho_{F\text{-}DCM}=0$}
         \label{Fig:pij2007-b}
     \end{subfigure}
     \begin{subfigure}[h]{0.75\textwidth}
         \centering
         \includegraphics[width=\textwidth]{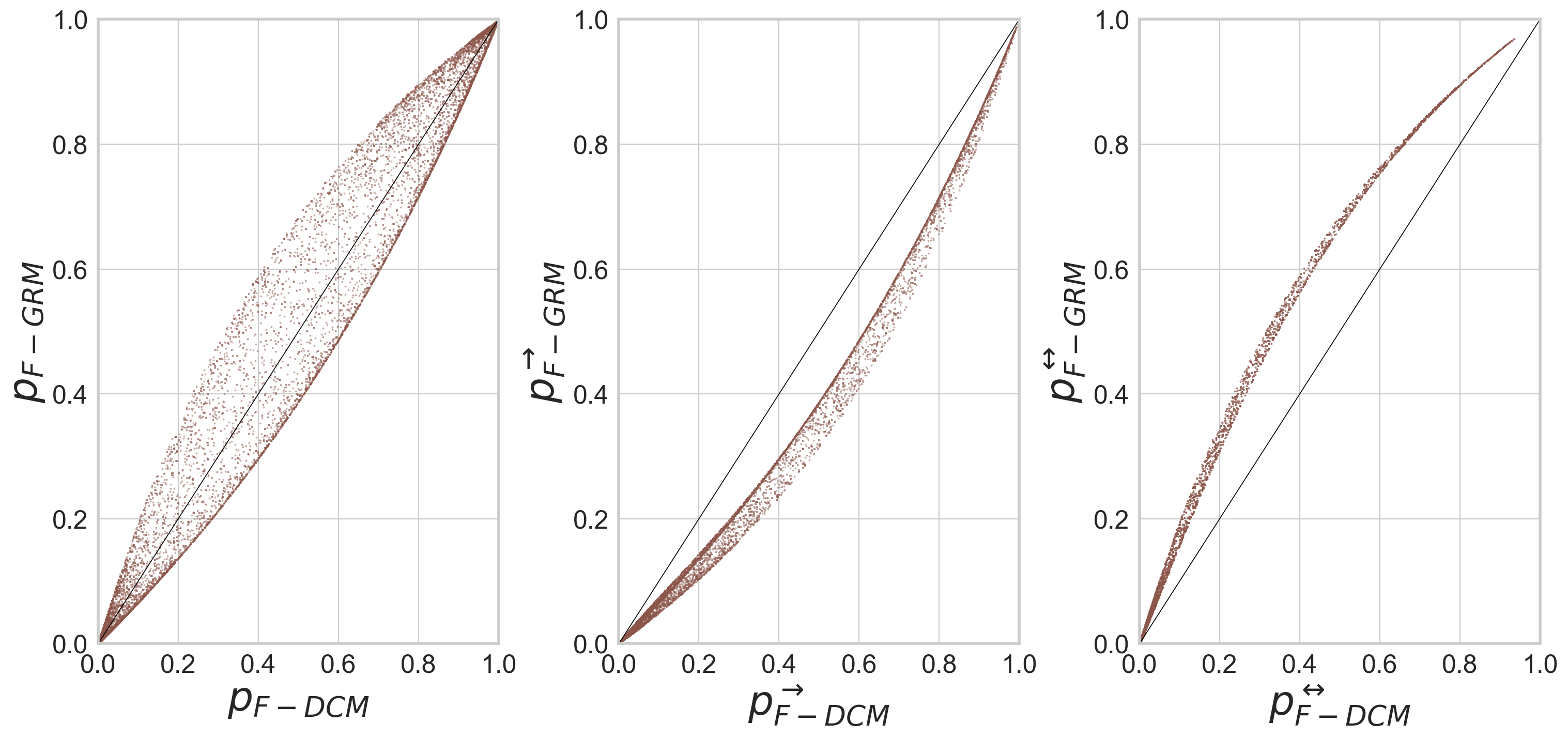}
         \caption{$\Delta_t=245$ days, $\rho_{F\text{-}DCM}=-0.19$}
         \label{Fig:pij2007-c}
     \end{subfigure}
     
        \caption{Link probability, 2007. Comparison between F-DCM and F-GRM. On the left there are the unconditional probabilities, on the center there are the mono-directed probabilities and on the right, there are the bi-directed probabilities. Different colors correspond to different aggregation periods $\Delta_t$: in red (top) and brown (bottom) there are the cases of the minimum and maximum value of $\rho_{F\text{-}DCM}$ while in purple (center) there is the case of $\rho_{F\text{-}DCM}\sim0$.}
        \label{Fig:pij_2007}
\end{figure}

In Figs. \ref{Fig:pij_2007}, we present a comparison of link probabilities between F-DCM and our model F-GRM, considering three different aggregation periods, in 2007. 
\subsection{z-score}
\begin{figure}[h]
    \centering
    \includegraphics[width=1\textwidth]{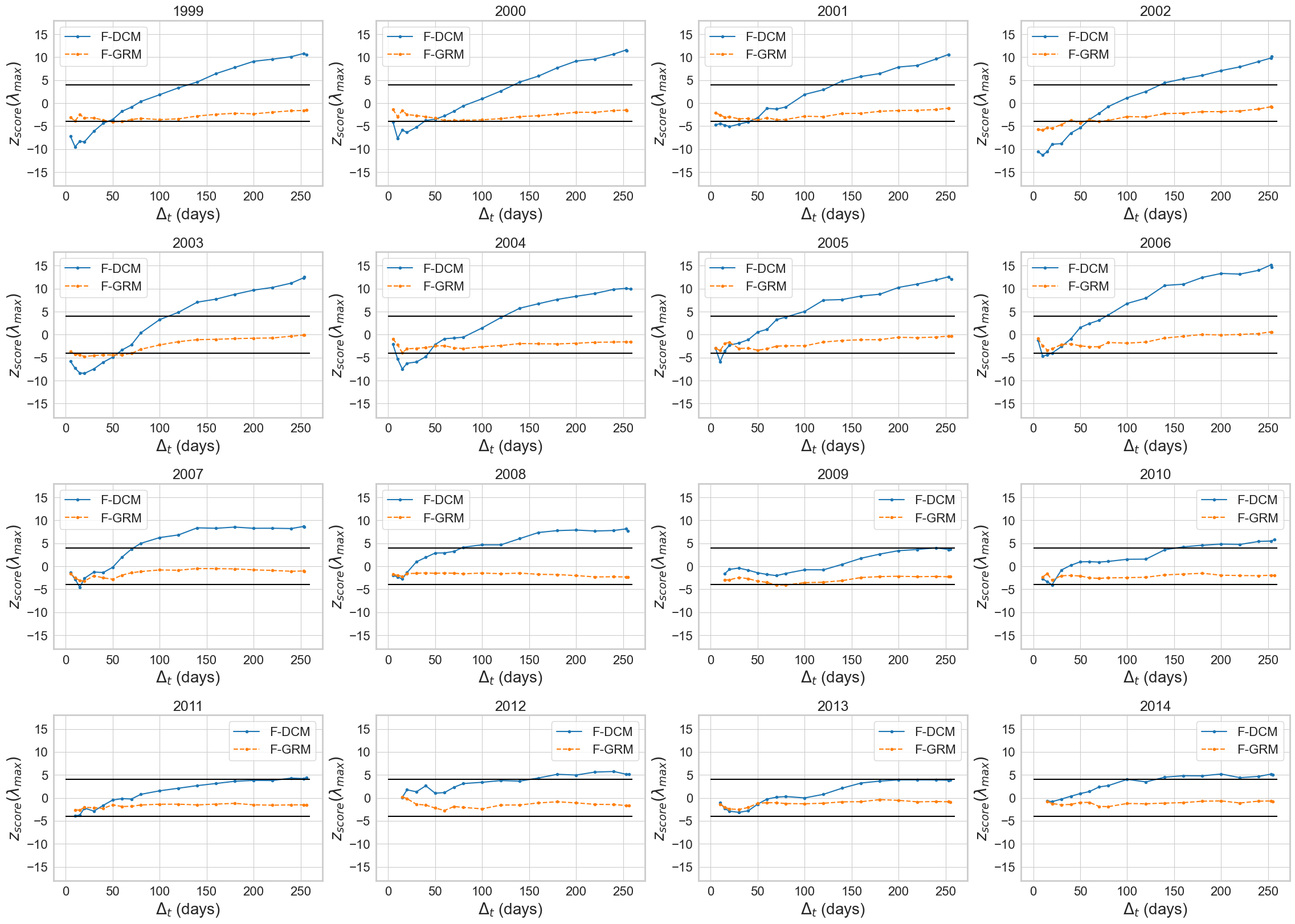}
    \caption{$z_{score}$ of the maximum eigenvalue as the aggregation period increases, in the period 1999-2014. Horizontal black lines show the values $z_{score}\pm 4$. F-DCM in blue and F-GRM in orange.}
    \label{z_score_appendix}
\end{figure}

Fig. \ref{z_score_appendix} presents the $z_{score}$ of the leading eigenvalue in the years 1999-2014, defined in Eq. \ref{eq:z-score}.
\subsection{Histogram of $\tau$}
\begin{figure}[h]
    \centering
    \includegraphics[width=0.6\textwidth]{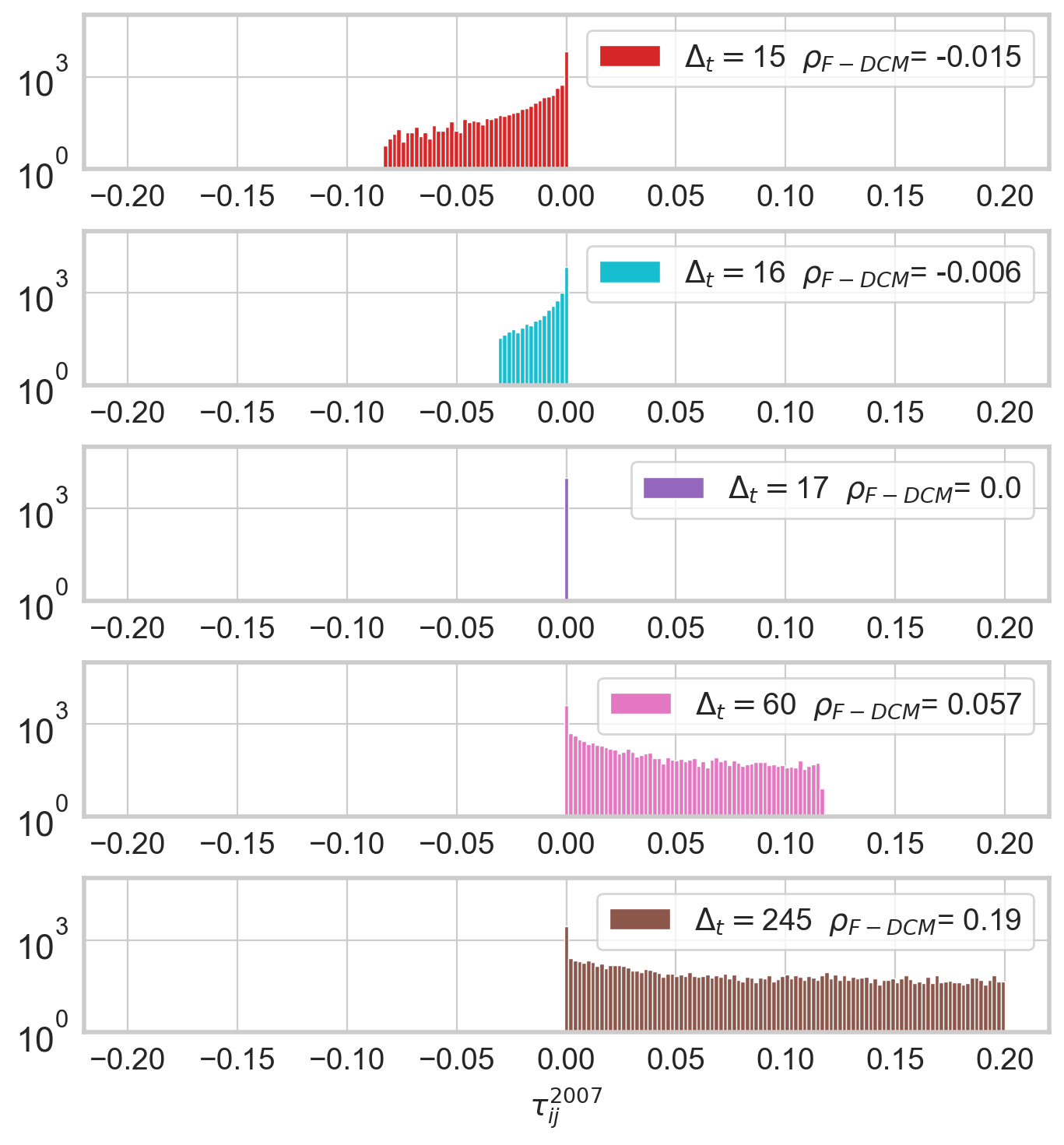}
    \caption{Histogram of $\tau_{ij}$, 2007. Different colors correspond to different aggregation periods $\Delta_t$ (days): in red and brown there are the cases of the minimum and maximum value of $\rho_{F\text{-}DCM}$ while in purple there is the case of $\rho_{F\text{-}DCM}\sim0$.}
    \label{Fig:hist_tau_2007}
\end{figure}

Fig. \ref{Fig:hist_tau_2007} shows the distribution of the correlation $\tau_{ij}$; different colors correspond to different aggregation periods in 2007.
\subsection{Bulk of the spectra}
\begin{figure}[h]
    \centering
    \includegraphics[width=0.9\textwidth]{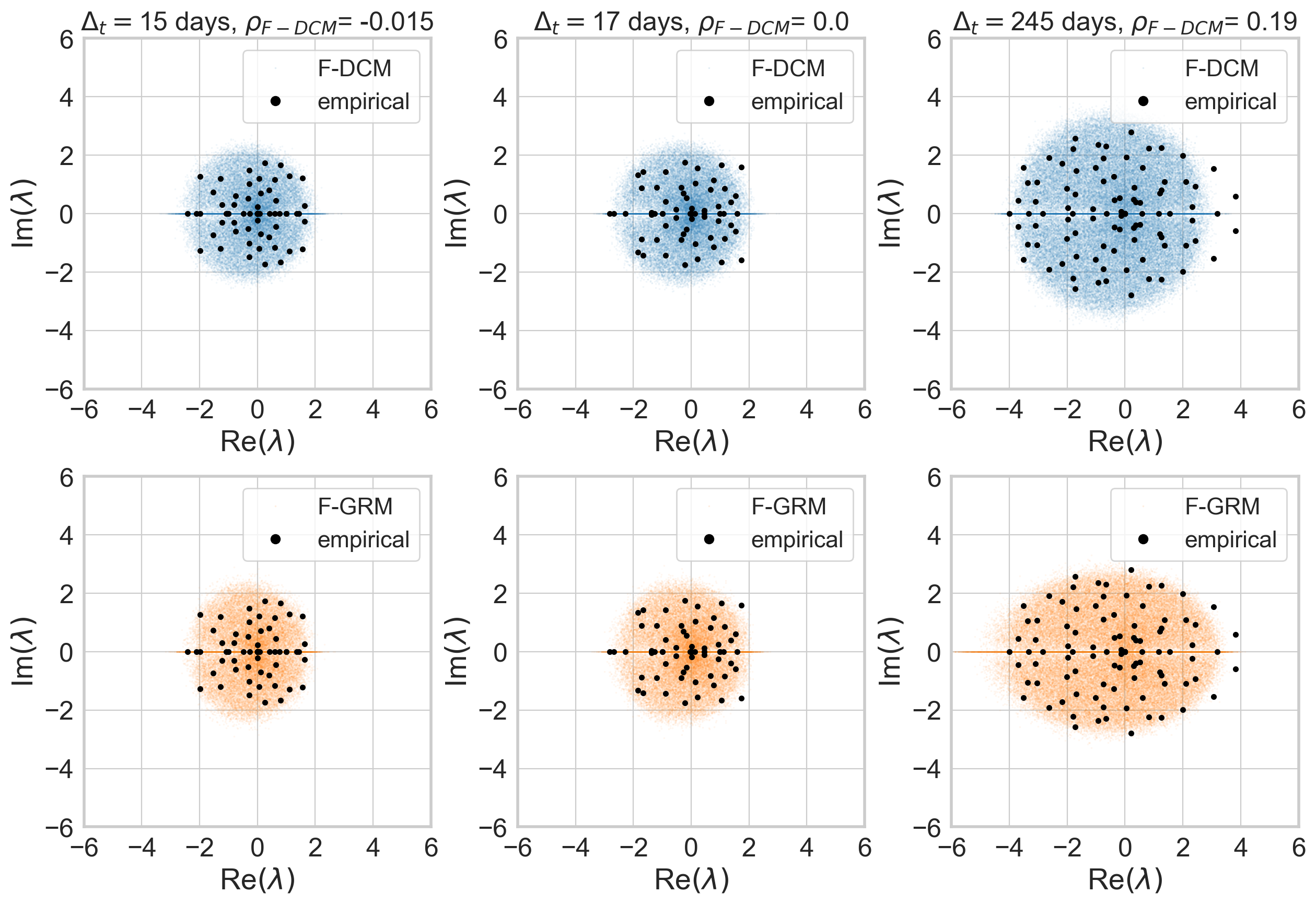}
    \caption{Bulk of the spectra of the adjacency matrix, 2007, different aggregation periods: whose $\rho_{F-DCM}$ is the yearly minimum, zero and maximum. Comparison between F-DCM in blue, and F-GRM in orange.}
    \label{Fig:ellipseaij_2007}
\end{figure}

Fig. \ref{Fig:ellipseaij_2007} reports the bulk of the spectra of the ensemble of the generated adjacency matrices in 2007; F-DCM in blue and F-GRM in orange. Black dots represent the eigenvalues of the empirical network. From the left to the right, there are reported aggregation periods that correspond to the minimum, zero and maximum value of $\rho_{F-DCM}$, respectively.
\clearpage
\section{Hypothesis of elliptical spectra}\label{appendix:ellipse}
In \cite{sommers1988spectrum} it was found that the eigenvalue density for a square real asymmetric matrix $J$, under assumptions specified below, is uniform over an ellipse in the complex plane, whose real and imaginary axes are $1+\tau$ and $1 - \tau$, respectively. The assumptions are:
\begin{equation}
    \braket{J_{ij}}=0 
\end{equation}
\begin{equation}
    \braket{J_{ij}^2}=1/N
\end{equation}
\begin{equation}
    \braket{J_{ij}J_{ji}}=\tau/N
\end{equation}
We consider the adjacency matrix A.
In order to obtain zero mean we should impose $J^0_{ij}=A_{ij}-p_{ij}$.
Then,
\begin{equation}
    \braket{(J^0_{ij})^2}=\braket{(a_{ij}-p_{ij})^2}=p_{ij}(1-p_{ij})
\end{equation}
\begin{equation}
    \braket{J^0_{ij}J^0_{ji}}=\braket{(a_{ij}-p_{ij})(a_{ji}-p_{ji})}=p^\leftrightarrow_{ij}-p_{ij}p_{ji}
\end{equation}

In order to also obtain $\braket{J_{ij}^2}=1/N$, we consider $J^1_{ij}=\frac{A_{ij}-p_{ij}}{\sqrt{Np_{ij}(1-p_{ij})}}$

Then:
\begin{equation}
    \braket{J^1_{ij}J^1_{ji}}=\frac{1}{N}\braket{\frac{A_{ij}-p_{ij}}{\sqrt{p_{ij}(1-p_{ij})}}\frac{A_{ji}-p_{ji}}{\sqrt{p_{ji}(1-p_{ji})}}}=\frac{p^\leftrightarrow_{ij}-p_{ij}p_{ji}}{N\sqrt{p_{ij(1-p_{ij})p_{ji}(1-p_{ji})}}}
\end{equation}

\subsection{GRM model}
We consider a simpler functional form than our model (only one free parameter):
\begin{equation}
p_{ij}^\rightarrow=\frac{x_i y_j}{1+x_i y_j+x_j y_i+ v^2 x_i y_j x_j  y_i}
\label{eq:pmono_grm_appendix}
\end{equation}
\begin{equation}
p_{ij}^\leftrightarrow=\frac{v^2  x_i y_j x_j  y_i}{1+ x_i y_j+ x_j y_i+v^2  
x_i y_j x_j  y_i}
\label{eq:pbi_grm_appendix}
\end{equation}
\begin{equation}
    p_{ij}=p_{ij}^\leftrightarrow+p_{ij}^\rightarrow.
\end{equation}

We define the denominator $w_{ij}=1+ x_i y_j+ x_j y_i+v^2  x_i y_j x_j  y_i$.

We have:
\begin{dmath}
%\begin{split}
  p^\leftrightarrow_{ij}-p_{ij}p_{ji}= \frac{x_ix_jy_iy_j}{w_{ij}^2} [v^2w_{ij}-(1+v^2x_iy_j+v^2x_jy_i+v^4 x_iy_jx_jy_i)]\\
  = \frac{x_ix_jy_iy_j}{w_{ij}^2} [v^2+v^2x_iy_j+v^2x_jy_i+v^4x_ix_jy_iy_j-1-v^2x_iy_j-v^2x_jy_i-v^4x_iy_jx_jy_iy]\\
  =\frac{x_ix_jy_iy_j}{w_{ij}^2} [v^2-1],
%\end{split}
\end{dmath}

\begin{equation}
    p_{ij}(1-p_{ij})=\frac{x_iy_j(1+v^2x_jy_i)(1+x_jy_i)}{w^2_{ij}},
\end{equation}

\begin{dmath}
  p_{ij}(1-p_{ij})p_{ji}(1-p_{ji})= \frac{x_ix_jy_iy_j}{w_{ij}^4}
  [(1+v^2x_jy_i)(1+v^2x_iy_j)(1+x_jy_i)(1+x_iy_j)]\\
  = \frac{x_ix_jy_iy_j}{w_{ij}^4}[1+(v^2+1)(x_iy_j)+(v^2+1)(x_jy_i)+(v^2+1)^2(x_ix_jy_iy_j)+v^2x_i^2y_j^2+v^2x_j^2y_i^2+v^2(v^2+1)(x_iy_j)(x_ix_jy_iy_j)+v^2(v^2+1)(x_jy_i)(x_ix_jy_iy_j)+v^4x_i^2x_j^2y_i^2y_j^2]\\= \frac{x_ix_jy_iy_j}{w_{ij}^4}f_{ij}^2,
\end{dmath}
where
\begin{dmath}
    f_{ij}^2=1+(v^2+1)[x_iy_j+x_jy_i]+(v^2+1)^2x_ix_jy_iy_j+v^2x_i^2y_j^2+v^2x_j^2y_i^2+v^2(v^2+1)(x_ix_jy_iy_j)[x_iy_j+x_jy_i]+v^4x_i^2x_j^2y_i^2y_j^2.
\end{dmath}

Then:
\begin{equation}
\begin{split}
    \braket{J^1_{ij}J^1_{ji}}&=(v^2-1)\frac{\sqrt{x_ix_jy_iy_j}}{N f_{ij}}.
\end{split}
\end{equation}

Recall that:
\begin{dmath}    w_{ij}^2=1+2x_iy_j+2x_jy_i+2(v^2+1)x_ix_jy_iy_j+x_i^2y_j^2+x_j^2y_i^2+2v^2(x_iy_j)(x_ix_jy_iy_j)+2v^2(x_jy_i)(x_ix_jy_iy_j)+v^4x_i^2x_j^2y_i^2y_j^2
\end{dmath}
so in the case $v^2=1$, we have $f_{ij}=w_{ij}$.
To reconcile our notation with \cite{sommers1988spectrum},
\begin{equation}
\begin{split}
    \braket{J^1_{ij}J^1_{ji}}&=\frac{1}{N}(v^2-1)\frac{\sqrt{x_ix_jy_iy_j}}{f_{ij}}=\frac{\tau}{N}.
\end{split}
\end{equation}
In the case $v^2=1$ the numerator is zero so $\braket{J^1_{ij}J^1_{ji}}=0$. In this case, where $p_{ij}^\leftrightarrow=p_{ij}p_{ji}$ by comparing with the notation in \cite{sommers1988spectrum}, $\tau=0$ that corresponds to the fully asymmetric ensemble in which $J_{ij}$ and $J_{ji}$ are independent. In this case, the bulk of the spectra is a circle.

If we reintroduce the missing free parameter $u$ that tunes the density, we have the following transformation:
\begin{equation}
    x_iy_j\rightarrow u\mathcal{A}_i\mathcal{L}_j
\end{equation}
where $\mathcal{A}, \mathcal{L}$ are respectively the total interbank assets and liabilities.
We define
\begin{dmath}
    g_{ij}^2=1+u(v^2+1)[\mathcal{A}_i\mathcal{L}_j+\mathcal{A}_j\mathcal{L}_i]+u^2(v^2+1)^2\mathcal{A}_i\mathcal{A}_j\mathcal{L}_i\mathcal{L}_j+u^2v^2\mathcal{A}_i^2\mathcal{L}_j^2+u^2v^2\mathcal{A}_j^2\mathcal{L}_i^2+u^3v^2(v^2+1)(\mathcal{A}_i\mathcal{A}_j\mathcal{L}_i\mathcal{L}_j)[\mathcal{A}_i\mathcal{L}_j+\mathcal{A}_j\mathcal{L}_i]+u^4v^4\mathcal{A}_i^2\mathcal{A}_j^2\mathcal{L}_i^2\mathcal{L}_j^2
\end{dmath}
thus
\begin{equation}
\begin{split}
    \braket{J^1_{ij}J^1_{ji}}&=\frac{1}{N}u(v^2-1)\frac{\sqrt{\mathcal{A}_i\mathcal{A}_j\mathcal{L}_i\mathcal{L}_j}}{g_{ij}}=\frac{\tau_{ij}}{N}.
\end{split}
\end{equation}

\subsection{F-DCM model}
For F-DCM, since $\braket{(A_{ij}-p_{ij})(A_{ji}-p_{ji})}=0$, it is $\tau=0$. As a consequence, the bulk of the spectra is always a circle.
\clearpage

 \bibliographystyle{elsarticle-num}
 \bibliography{ref}





% \end{thebibliography}
\end{document}